\documentclass[%
 aip,
 amsmath,amssymb,
 reprint,%
]{revtex4-1}

\usepackage{graphicx}
\usepackage{dcolumn}
\usepackage{bm}

\usepackage[utf8]{inputenc}
\usepackage[T1]{fontenc}
\usepackage{mathptmx}
\usepackage{etoolbox}
\usepackage{graphicx}
\usepackage{float}
\usepackage{subfig}
\usepackage{amsmath}
\usepackage{amssymb}

\makeatletter
\def\@email#1#2{%
 \endgroup
 \patchcmd{\titleblock@produce}
  {\frontmatter@RRAPformat}
  {\frontmatter@RRAPformat{\produce@RRAP{*#1\href{mailto:#2}{#2}}}\frontmatter@RRAPformat}
  {}{}
}%
\makeatother
\begin{document}

\preprint{AIP/123-QED}

\title{An adaptive primitive-conservative scheme for high speed transcritical flow with an arbitrary equation of state}

\author{Bonan Xu }
\affiliation {School of Aeronautics and Astronautics, Zhejiang University, Hangzhou 310027, PR China.}
\author{Hanhui Jin}
\affiliation {School of Aeronautics and Astronautics, Zhejiang University, Hangzhou 310027, PR China.}
\affiliation {State Key Laboratory of Clean Energy Utilization, Zhejiang University, Hangzhou, 310027, PR China}
\email{enejhh@emb.zju.edu.cn.}
\author{Yu guo}
\affiliation {School of Aeronautics and Astronautics, Zhejiang University, Hangzhou 310027, PR China.}
\author{Jianren Fan}
\affiliation {State Key Laboratory of Clean Energy Utilization, Zhejiang University, Hangzhou, 310027, PR China}

\date{\today}

\begin{abstract}
When fully conservative methods are used to simulate transcritical flow, spurious pressure oscillations and numerical instability are generated. The strength and speed of propagation of shock waves cannot be represented correctly using a semi-conservative or primitive method. In this research, an adaptive primitive-conservative scheme is designed to overcome the aforesaid two difficulties. The underlying cause for pressure oscillation is analyzed within the framework of Finite Volume Method (FVM). We found that the nonlinearity of the thermodynamic properties of transcritical fluids renders standard conservative numerical methods ineffective. In smooth regions, schemes based on primitive variable are used to eliminate spurious pressure oscillations. For the purpose of correctly capturing shock waves, the modified Roe Riemann solver for real fluid is utilized in regions where shock waves induce discontinuity. The adaptive numerical approach relies only on the speed of sound, eliminating the requirement to calculate the derivatives of thermodynamic quantities. A large number of numerical test cases conducted in one- and two-dimensional spaces have shown the robustness and accuracy of the proposed adaptive scheme for the simulations of high speed transcritical flows. 
\end{abstract}

\maketitle

\section{\label{sec:intro}  Introduction}

\begin{figure}[h]
	\centering
	\subfloat[Density] {\includegraphics[scale=0.35]{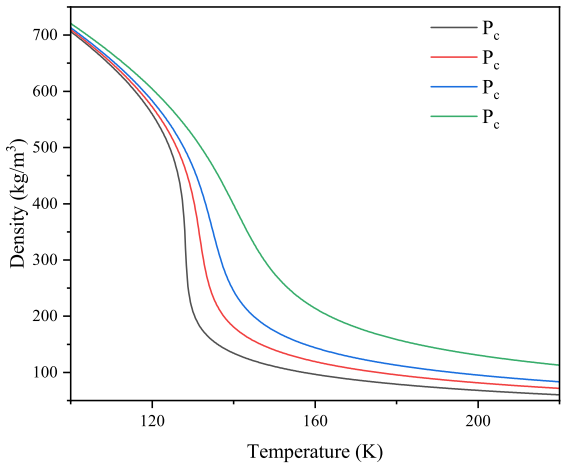}}
	\quad
	\subfloat[Constant pressure specific heat capacity]{\includegraphics[scale=0.35]{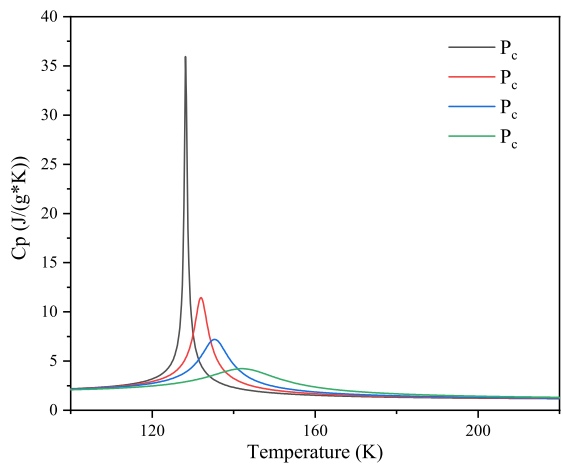}}
	\caption{Density and constant pressure specific heat capacity of transcritical nitrogen at different pressures. The critical pressure $P_c$ of nitrogen is $ 3.3958 MPa $ and the critical temperature $T_c$ is 126.192 K}
	\label{fig::fig1}
\end{figure}

Understanding the mechanics of interaction between transcritical layer and shock waves is essential for a variety of industrial applications, including high-speed fuel injections in advanced internal combustion engines\cite{internal1, internal2, internal3} and gas turbine\cite{turbine1, turbine2,turbine3}. 	In Fig.\ref{fig::fig1} we illustrate the density and constant pressure specific heat capacity of transcritical nitrogen.These data are downloaded from National Institute of Standards and Technology (NIST).The thermodynamic and transport properties of fluids exhibit a high degree non-linearity and huge gradient in the region near Widom line \cite{widomline}. This nonlinearity reduces as pressure increases. In transcritical process, when the ambient pressure exceeds the critical value of fluids, the injected low temperature fluids are heated from liquid-like state to gas-like state in a thin layer and vice versa. The transcritical process with the properties of fluid dramatically changing has not been completely understood\cite{MDdrop, interface} and numerous novel occurrences have been reported in recent years. For example, analytical investigations\cite{analytical}, experimental observations\cite{experimental}, and numerical simulations\cite{MD} have shown that surface tension, which is thought to approach zero under supercritical pressure, nevertheless exists in the transcritical regions. When shock waves interact with the thin transcritical layers with significant nonlinearity, a more complicated structure will be formed. Numerical simulations may provide much more details for a deeper understanding about the mechanism of transcritical flows. To effectively simulate this circumstance, numerical approaches must be robust in the transcritical zone and accurately capture the shock waves, which poses a significant challenge to standard numerical algorithms. As a result, certain new numerical methods need be developed to get a better knowledge of high speed transcritical flows. There are two crucial aspects to consider while constructing a numerical approach for the simulation of the interactions between transcritical layers and shock waves. To begin, the numerical approach must be robust in the transcritical region. Second, shock wave strength and propagation speed can be appropriately resolved.

Unfortunately, due to the high no-linearity and dramatically change in thermodynamic properties of fluids in transcritical region, spurious pressure oscillations\cite{MA_DF, drop_shock, osc_pre_sp, thermoacoustic} will be introduced while solving the conservative form Euler equations using the conservative schemes. This spurious pressure oscillation will produce a significant oscillation in velocity, leading numerical method to fail. The fundamental cause of pressure oscillations when resolving transcritical fluids will be analytically and numerically investigated in the context of the Finite Volume Method (FVM). Even in ideal gas flow, the pressure oscillations may also be observed in the regions that thermodynamic properties quickly change, such as material interfaces\cite{material_interfaces} or multicomponent flows\cite{Multicomponent1}. Numerous attempts have been made to remove these non-physical pressure oscillations in order to create a simulation that is accurate and robust. Using a low-Mach-number approximation of the governing equations, Pasquale\cite{LMa1} explored a low-velocity transcritical nitrogen jet with direct numerical simulation in order to completely resolve the interaction between the turbulent and transcritical layers. The low-Mach number approximation can yield accurate results for low-velocity fluids, but it generates non-negligible error when the flow velocity exceeds 0.1 Mach. The double-flux method\cite{DB_OR}, which was originally developed to eliminate pressure oscillations in compressible multicomponent fluids, was improved and expanded to simulate transcritical flows\cite{MA_DF}. From then on, many new numerical method\cite{MA_DF, DB_HB, PC-SAFT} for transcritical flows were developed based on double-flux method. The assumption of constant specific heat ratio and internal energy in the double-flux technique eliminates pressure oscillations in the transcritical region, but at the expense of energy conservation and a quicker prediction of shock speed\cite{MA_DF}. Bradley Boyd and Dorrin Jarrahbashi\cite{DB_HB} proposed a hybrid method that switches between double-flux and traditional fully-conservative numerical methods to address the aforementioned two issues. Instead of total energy equation, discrete pressure evolution equation was solved\cite{press} using high-order compact differencing scheme, resulting in robust and accuracy simulations of gas–liquid-like flows under supercritical pressures. However, the energy conservation equation can only be transformed to the pressure evolution equation in the smooth, shock-free region, limiting the applicability of these kind of numerical methods to subsonic flows only. In addition, various simulation methods\cite{vanserw, ader} for flows with a nonlinear Equation of State (EoS) have been developed in recent years. However, these approaches are often restricted to some specific EoS.

In high-velocity transcritical flows, resolving the strength and propagation speed of velocity is an additional significant challenge. The work of Hou and Le Floch\cite{math_fund} demonstrates that, similar to conservative methods, non-conservative methods converge to the proper solution for shock-free smooth flow, hence the pressure evolution equation can be used to maintain pressure equilibrium in smooth flows. On the other aspect, they found that non-conservative schemes cannot converge to the proper solution when the shock waves exist but can be locally corrected by adopting conservative scheme. Their results provide a solid theoretical basis for the development of adaptive conservative-primitive scheme. Thus, in order to maintain the robustness and correction, the numerical methods adapting from conservative forms to primitive forms have been used in many fields\cite{ad_toro, ad_osher, ad1} including multiphase flows, flows with nonlinear EOS. In this paper, we introduce an adaptive method for the simulation of high-speed transcritical flows. A shock sensor based on the Primitive Variable Riemann Solver(PVRS)\cite{toro2013riemann} would be used to identify the cells involved with shock waves. The Euler equation of primitive form is employed in the shock free smooth region whereas the cells near shock waves are updated based on the conservative form.

In transcritical flows, the ideal gas EoS cannot describe the high non-linearity of real fluid, hence it is essential to describe the thermodynamic and transport properties of real fluids using more complicated EoS. The EoS mainly influences the numerical methods in the following two aspects. First, the nonlinearity and abrupt change of thermodynamic and transport quantities in the transcritical region must be accurately described. For single-species flows, several researches\cite{press, LMa1} directly used the NIST look-up table approach to get precise value of fluid properties. Obtaining the partial derivatives of thermodynamic variables using the NIST lookup tables might be difficult and computationally costly. In addition, it is hard to apply this approach to multicomponent flows due to the fact that the mixing rule for real gas cannot be applied to real gas. The cubic EoS, including SRK\cite{SRK} (Soave–Redlich–Kwong) and PR (Peng–Robinson)\cite{PR} EoS, achieve a good balance between accuracy and computational cost. In industrial applications, cubic EoS are often used\cite{UC1} to represent real fluid properties in simulations under transcritical or supercritical conditions. For a more precise description of the thermophysical characteristics of real fluids, several researchers have used more complicated EoS, such as the Benedict–Webb–Rubin and Statistical Association Fluid Theory EoS\cite{PC-SAFT}. Second, the numerical methods themselves are affected by the properties of EoS. Even considering the popular Harten-Hyman entropy fix\cite{entropy_fix}, the results of the Roe method will yield entropy violating discontinuities in strong rarefaction waves when employing the cubic EoS. Previous studies have shown that the fluxes of the Euler equation are convex, hence this discontinuity should not exist in rarefaction. Wang et.al\cite{StARS, expansion} recently suggested a structurally complete approximate Riemann solution (StARS) for transcritical flow in the context of cubic state equations in order to restore the expansion wave. Some complex EoS may affect the convexity of isentropes\cite{noncv_EOS, muller1999riemann,real_Roe}, resulting in anomalous wave structures such as composite or split waves. In this paper, the proposed adaptive primitive-conservative scheme can be employed with arbitrary EoS, but only under the assumption that the isentrope is convex. We note that the convexity of the isentropes of fluids and their components in the transcritical regime remains an open question, which is outside the scope of this research.

This paper is organized as follows: In section 2, the governing equations are presented in both conservative and primitive forms. The EoS utilized in this work for evaluating the thermodynamic properties of transcritical fluids is discussed in Section 3. Then, we analyze the underlying reason for the failure of the conservative methods while addressing transcritical flows in the context of the FVM, which may serve as a guideline for the future design of numerical methods for flows with high non-linear properties. Without specifying a certain kind of EoS, we construct a novel adaptive primitive-conservative scheme for supersonic transcritical flow in Section 4. In Section 5, we evaluate the proposed numerical method using a variety of one-dimensional (1-D) and two-dimensional (2-D) numerical test cases. Comparing the numerical results with the available analytical data demonstrates that this adaptive scheme can accurately capture shock waves while removing oscillations. In section 6, we provide a concise conclusion.

\section{\label{sec:gov}  Governing equations}
\subsection{Conservative form}
The time-dependent Euler equations are of interest in this paper under the assumption that the flow is inviscid and isentropic. The conservative form of 1-D Euler equations can be written as:
\begin{equation}
	\frac{\partial \mathbf{U}}{\partial t} + \frac{\partial \mathbf{F}(\mathbf{U})}{\partial x} = 0
	\label{con_euler}
\end{equation}
Where the vector of conservative variables and fluxes $\mathbf{U}$ and $ \mathbf{F}\left(\mathbf{U}\right) $, are given as:
\begin{equation}
	\mathbf{U} = \begin{bmatrix} \rho \\ \rho u \\ \rho E  \end{bmatrix}, \qquad
	\mathbf{F} = \begin{bmatrix} \rho u \\ \rho u^2 + P \\ \rho u H \end{bmatrix}
	\label{con_flux}
\end{equation}
Here, $ \rho $, $ u $, $ p $ denote density, particle velocity and static pressure, while $ E $ and $ H $ imply total specific energy and total specific enthalpy, which are defined as follow:
\begin{equation}
	\begin{split}
		E &= \frac{1}{2}u^2 +e,  \\
		H &= E + \frac{P}{\rho} = \frac{1}{2}u^2 + e + \frac{P}{\rho}
		\label{enthalpy}
	\end{split}
\end{equation}
The specific internal energy $ e $ is determined by EoS utilized in simulations.

The conservative form of the Euler equations reflects the physical laws of mass conservation, momentum conservation, and energy conservation, and must be obeyed everywhere in the flow field, even across discontinuities, to ensure mass, momentum, and energy conservation.

\subsection{Primitive form}
The primitive form of time-dependent 1D Euler equations is:
\begin{equation}
	\frac{\partial \mathbf{W}}{\partial t} + \mathbf{B}\frac{\partial \mathbf{W}}{\partial x} = 0
	\label{pri_euler}
\end{equation}
where $ \mathbf{W} $ is the primitive vector and $ \mathbf{B} $ is the coefficient matrix of the primitive form of Euler equations, the expression is:
\begin{equation}
	\mathbf{W} = \begin{bmatrix} \rho \\ u \\ P \end{bmatrix},  \qquad 
	\mathbf{B} = \begin{bmatrix}u &\rho & 0 \\ 0 & u & \frac{1}{\rho} \\ 0 & \rho c^2 &u  \end{bmatrix}
\end{equation}
Here, $ c $ represents the sound speed that can be determined using the EoS. The above equations are derived at a partial-differential-equation level and independent to the types of EoS. The primitive form of Euler equations is derived through differentiable transformations of the conservative form and, assuming the flow field is smooth, is equivalent to the conservative form. As a result of the loss of the meaning of a classic derivative, the situation changes once the discontinuity forms. When employing the primitive form Euler equations, the strength and propagation speed of shock waves cannot converge to the right values. The adaptive primitive-conservative scheme is proven to converge to the proper solution while locally correcting the primitive form with the conservative form in the region of shock waves, as shown by earlier research\cite{ad_osher, ad1, ad_toro} and numerical data presented in this study.
\begin{figure}
	{\includegraphics[scale=0.55]{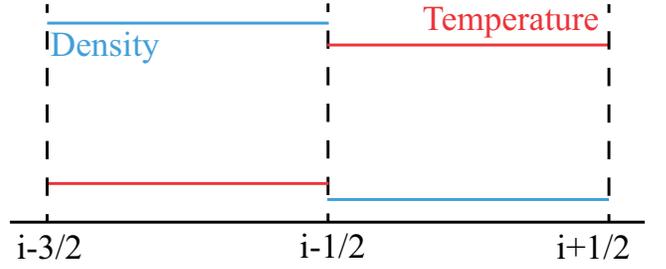}}
	\caption{Density and temperature discontinuities at the cell interface $ i-\frac{1}{2} $}
	\label{fig:interface}
\end{figure}
\begin{figure}
	{\includegraphics[scale=0.3]{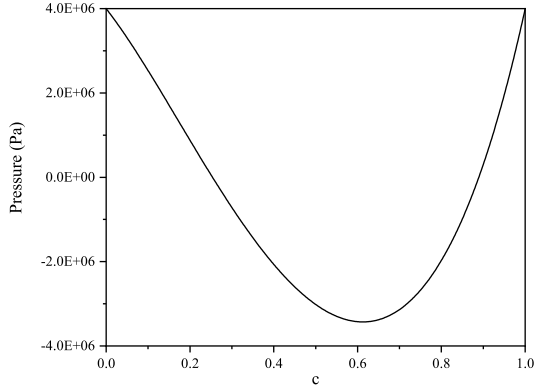}}
	\caption{The value of pressure in cell $ i $ at next step when solving conservative Euler equations with first order upwind Godunov scheme in the context of PR EoS}
	\label{fig:reason}
\end{figure}
\section{\label{sec::thermo} Thermodynamics}
\subsection{Equation of state}
For an appropriate description of the nonlinearity in the transcritical regime, it is necessary to use a more complex Eos or the tabulation approach. In this paper, the numerical method is developed without any assumption regarding the form of EoS. The PR EoS are utilized for computation, because it can accurately represent the thermodynamic properties of fluids around Widom line and easy for implementation. This EoS can be written as follows:
\begin{equation}
	P = \frac{\rho R_u T}{M_w - b \rho} + \frac{a \rho^2}{M_w^2 + 2M_w b \rho - b^2 \rho^2}
	\label{PR}
\end{equation}
where $R_u$ denotes the universal gas constant and $ T $ denotes the temperature. The parameters $ a $ and $ b $ are written as:
\begin{equation}
	a = 0.45724\left(  \frac{R_u^2 T_c^2}{P_c} \right), \quad b = 0.07780\left( \frac{R_u T_c}{P_c} \right)
\end{equation}

and are determined by critical temperature $ T_c $ and critical pressure $ P_c $. The selection of the EoS depends on the computational requirements for the precision of the thermodynamic parameters. The precision of the PR EoS is enough for the development of new numerical algorithms.

The specific internal energy $ e $ and speed of sound $ c $ can be determined based on the selected EoS. For a more detail discussion of cubic EoS and the computation of thermodynamics properties for real fluids, we recommend the paper published by Kim et al\cite{KIM}.

For the sake of completeness and simplicity, we only write the final forms of specific internal energy $ e $ and speed of sound $ c $ that are directly related to the scheme in this paper as:

\begin{equation}
	\begin{split}
		e(T, \rho) &= e_0(T) + \int_{\rho_0}^{\rho}\left[ \frac{P}{\rho^{2}} - \frac{T}{\rho^{2}} \left(  \frac{\partial P}{\partial T} \right)_{\rho}  \right]_{T} d\rho,  \\ 
		c^2 &= \left( \frac{\partial P}{\partial \rho} \right)_{s} = \frac{c_p}{c_v}\left( \frac{\partial P}{\partial \rho} \right)_T
	\end{split}
\end{equation}

Here, subscript $ 0 $ refers to the ideal gas state which can be represented by NASA polynomials. 

\subsection{Reason for pressure oscillations}
In this section, within the framework of FVM, we will detailly discuss the fundamental cause of the pressure oscillations when using conservative method with non-linear EoS. The discussion will be carried out in the context of 1-D transcritical advection case governed by Euler equations. For simplicity, we assume that the pressure and velocity in the flow field are constant. However, as seen in fig.\ref{fig:interface}  here is the density and temperature discontinuities at the cell interface $ i-\frac{1}{2} $. The following are the values of primitive variables on the left and right sides of this interface:
\begin{equation}
	\begin{split}
		\mathbf{U}_L &= \begin{bmatrix} \rho_l \\ u \\ E_l\end{bmatrix}  \quad \mathbf{U}_R = \begin{bmatrix} \rho_r \\ u \\ E_r\end{bmatrix}   \\
		&\text{and} \quad P_l = P_r = P
	\end{split}
\end{equation}
For this situation, the analytical answer is straightforward. The flow field is reconstructed to a piecewise constant state in each cell, and the FVM is employed.

The FVM is used with the flow field reconstructed to a piecewise constant state in each cell. By using the conservative first order Godunov method, the average values of density and specific internal energy at next time step are:
\begin{equation}
	\begin{split}
		\rho_{avg} &= c \rho_l + (1-c)\rho_r   \\
		e_{avg} &=\frac { c\rho_l e_l + (1-c) \rho_r e_r}{\rho_{avg}}
	\end{split}
\end{equation}
where the variable $ c $ is defined as $ c=\frac{u\Delta t}{\Delta x} $. The specific internal energy of ideal gas can be computed as follows:
\begin{equation}
	e = \frac{P}{(\gamma - 1)\rho}
	\label{eq::idea_energy}
\end{equation}

It is straightforward to prove that the pressure value obtained by solving $ e_{avg} $ and $ \rho_{avg} $ with the idea gas relationship is still a constant equal to the original value, which is consistent with the analytical solution. The correlations between thermodynamic characteristics, on the other hand, show a significant degree of nonlinearity in the transcritical region. In other words, the average values of conservative variables are thermodynamically inconsistent, resulting in a significant pressure oscillation when pressure is solved using conservative method with non-linear EoS.

To demonstrate this issue more clearly, we use a simple numerical test. Using PR EoS, we assume the density, temperature, and pressure values on the left and right sides of the interface are:
\begin{equation}
	\begin{split}
		\begin{bmatrix} \rho_l \\ T_l \\ P_l\end{bmatrix} &= \begin{bmatrix} 580.586 \ kg/m^3 \\ 120\ K \\ 4e6\ Pa \end{bmatrix} \\
		\begin{bmatrix} \rho_r \\ T_r \\ P_r\end{bmatrix} &= \begin{bmatrix} 74.7415 \ kg/m^3 \\ 200\ K \\ 4e6\ Pa \end{bmatrix}
	\end{split}
\end{equation}

The average values of energy $ e_{avg} $ and density $ \rho_{avg} $ are firstly obtained when solving Euler equations using the conservative method. EoS is then used to determine the average values of temperature $ T_{avg} $ and pressure $  P_{avg} $. Fig.\ref{fig:reason} illustrates the correlations between the value of pressure in cell $ i $ at next step when solving conservative Euler equations with first order upwind Godunov scheme in the context of PR EoS and the variable $ c $. The pressure oscillation will be produced immediately. 

Additionally, one can observe that the pressure oscillation does not occur if and only if the variable $ c $ equals $ 0 $ or $ 1 $. This is impossible for flow fields with a non-uniform distribution of velocity. Unfortunately, this kind of discontinuity in thermophysical properties, known as the contact discontinuity, is prevalent in high-velocity flows. In this study, the Euler equations in their primitive form are applied to the shock-free regions, while the shock-affected cells are locally corrected using the conservative form.
\begin{figure*}[h!t]
	\centering
	\subfloat[Density]{\includegraphics[scale=0.3]{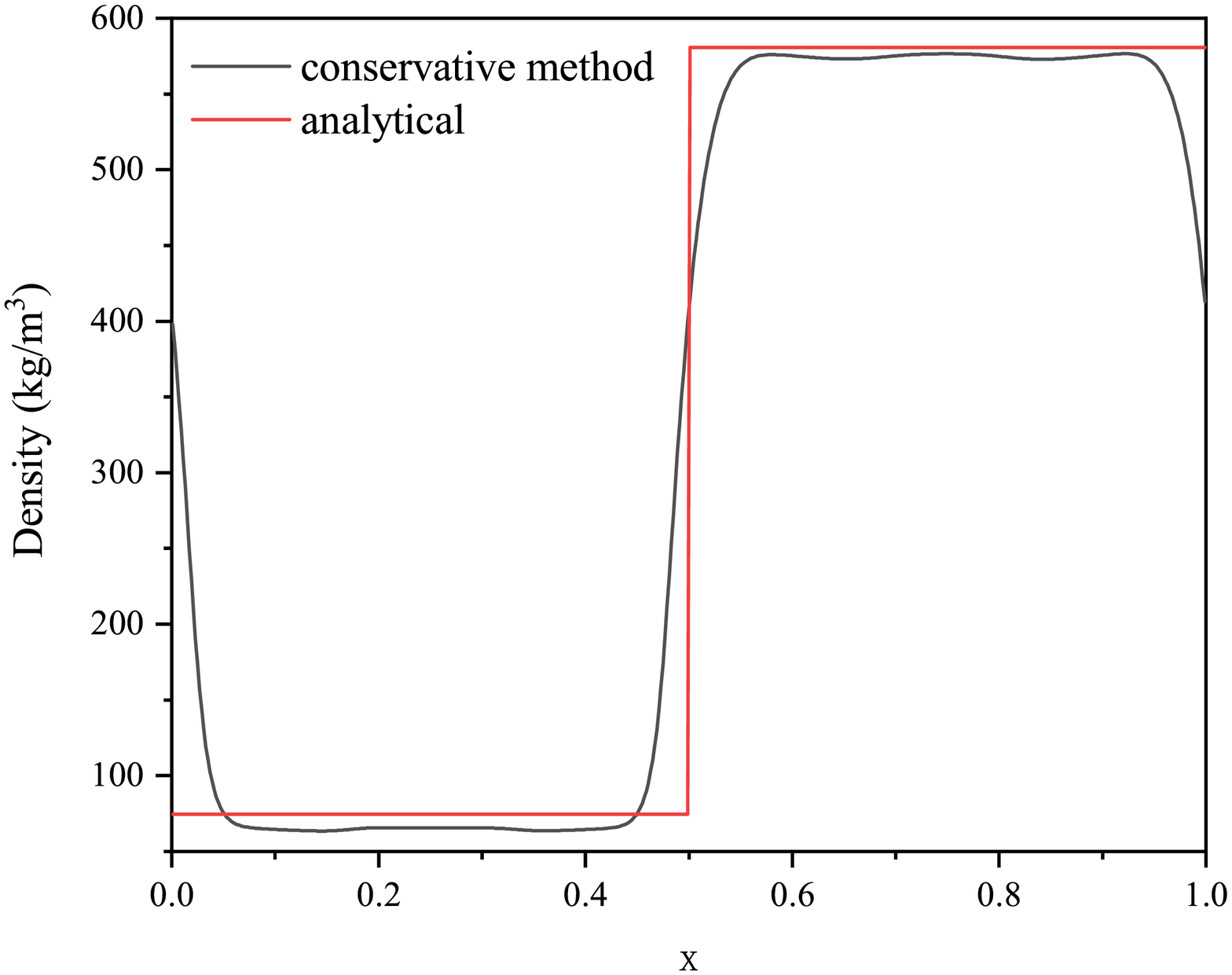}}
	\subfloat[Velocity]{\includegraphics[scale=0.3]{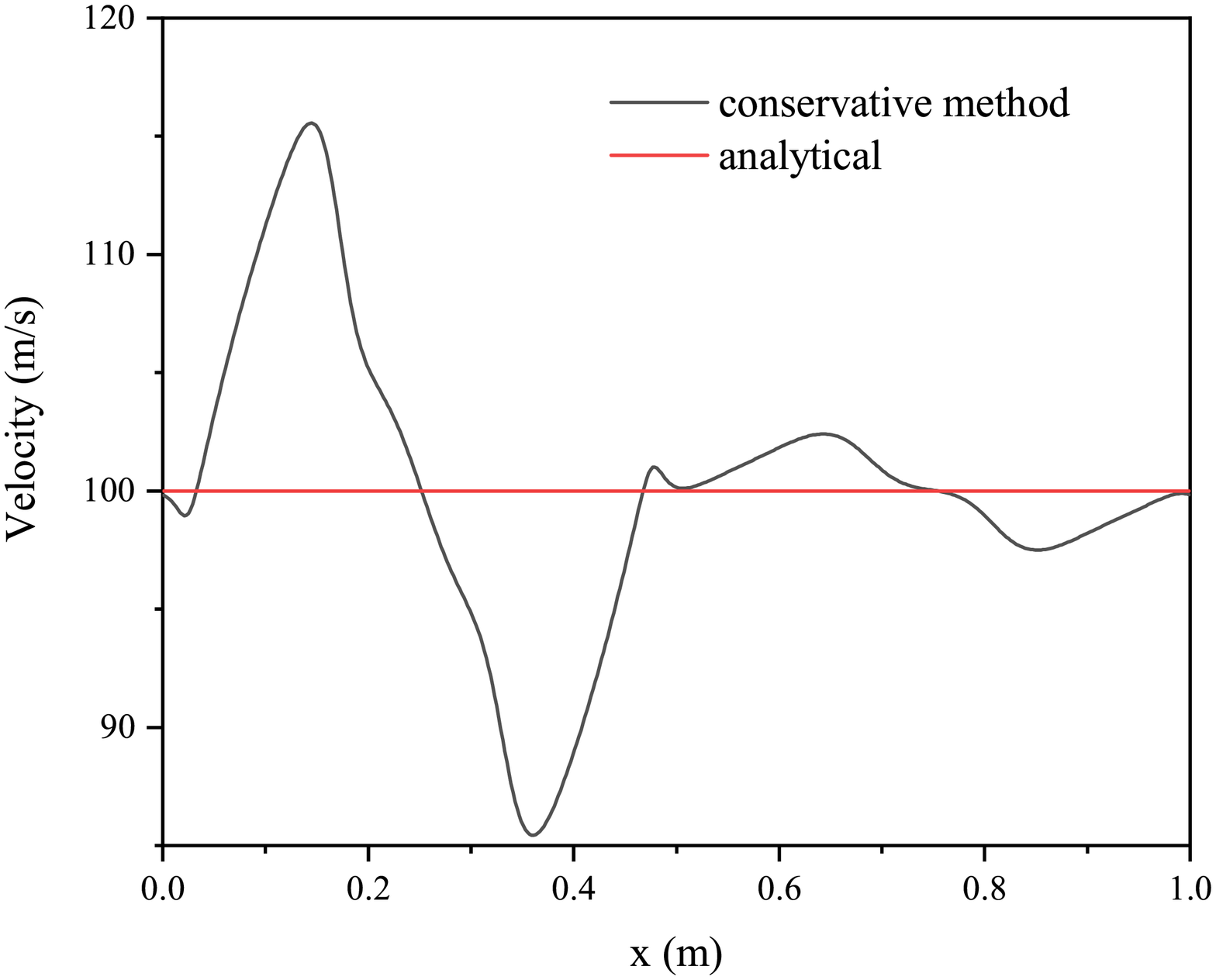}}
	\quad
	\subfloat[Pressure]{\includegraphics[scale = 0.3]{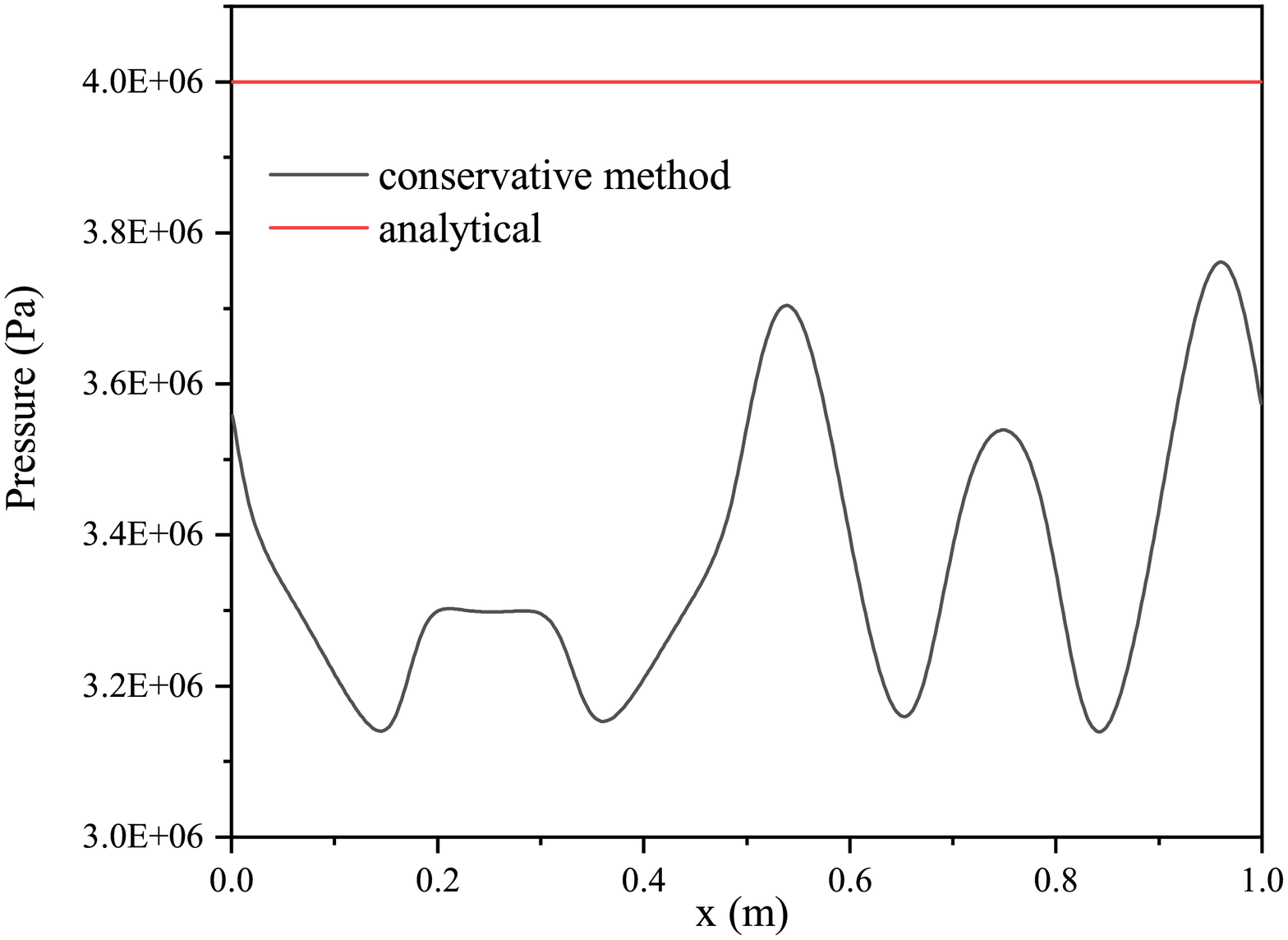}}
	\subfloat[Temperature]{\includegraphics[scale = 0.3]{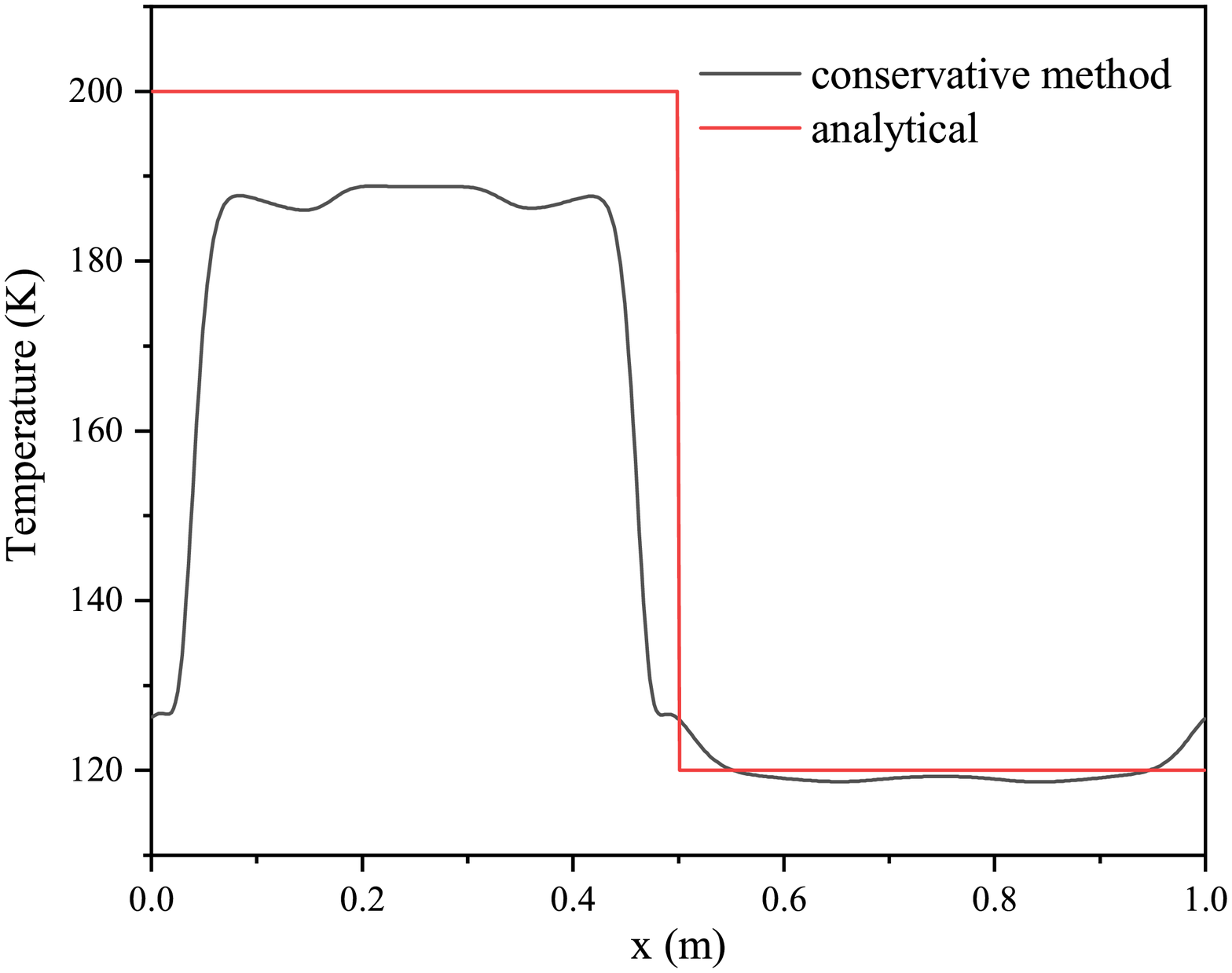}}
	\caption{The profiles of density, velocity, pressure and temperature when resolving 1-D transcritical advection case with sharp interface by conservative scheme}
	\label{fig::adcon}
\end{figure*}

\section{\label{sec::NM} Numerical method}

In this section, we will discuss the adaptive primitive-conservative hybrid scheme in detail. This scheme is primarily comprised of the three sub-modules listed below: Riemann solver for primitive Euler equations, Riemann solver for primitive Euler equations and its extension to real gas, and a sensor for shock waves.

\begin{figure*}[t]
	\centering
	\subfloat[Density]{\includegraphics[scale=0.3]{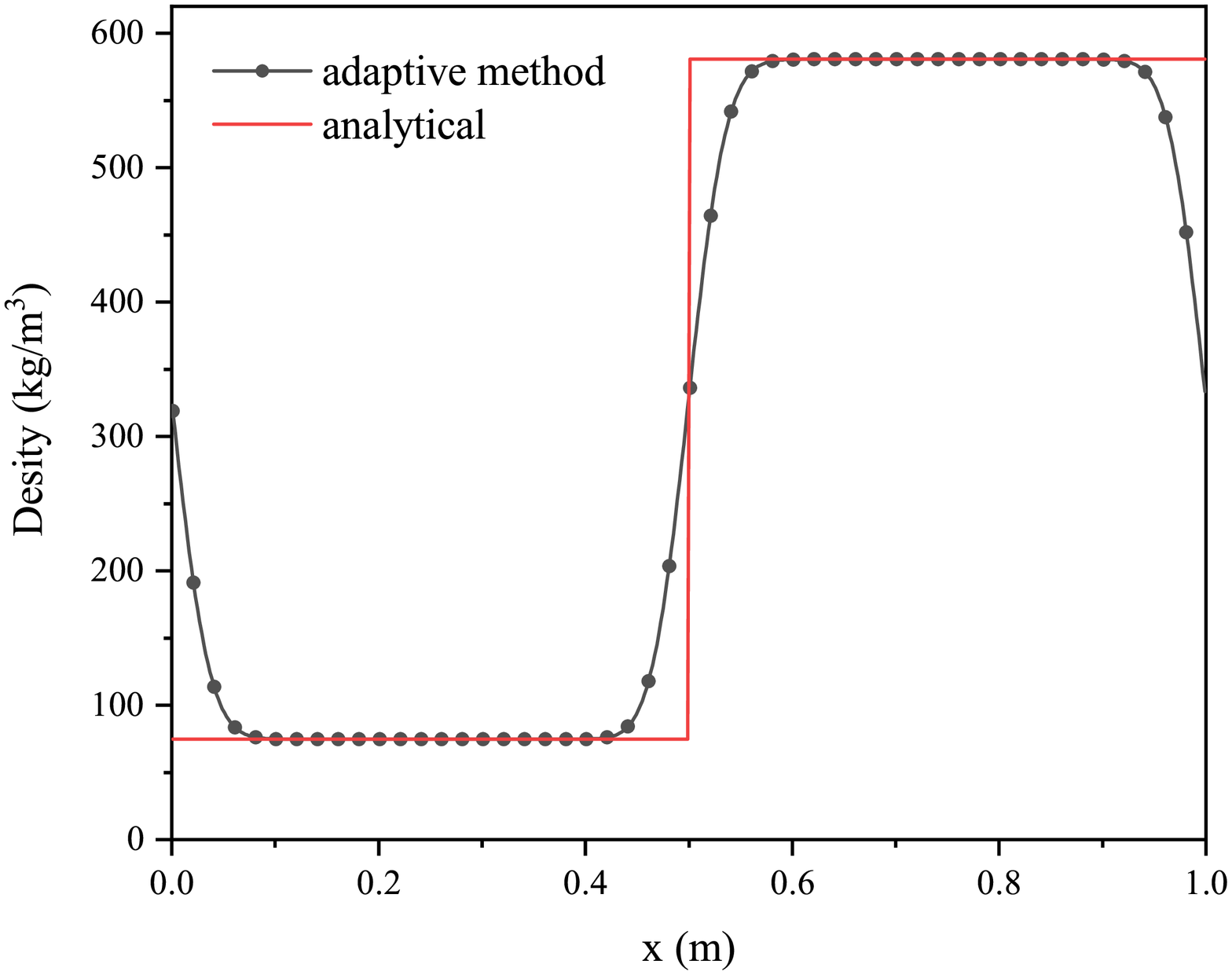}}
	\subfloat[Velocity]{\includegraphics[scale=0.3]{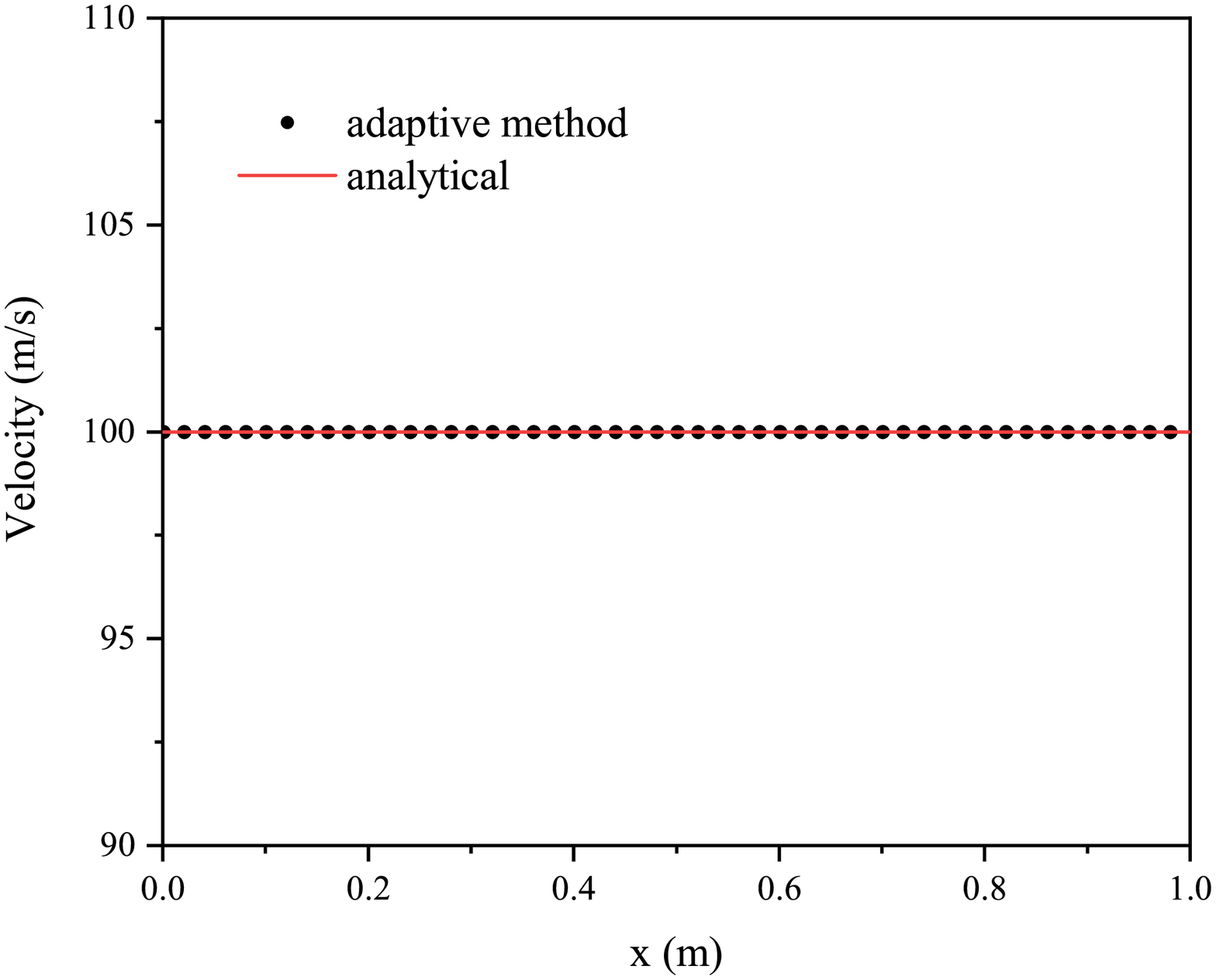}}
	\quad
	\subfloat[Pressure]{\includegraphics[scale = 0.3]{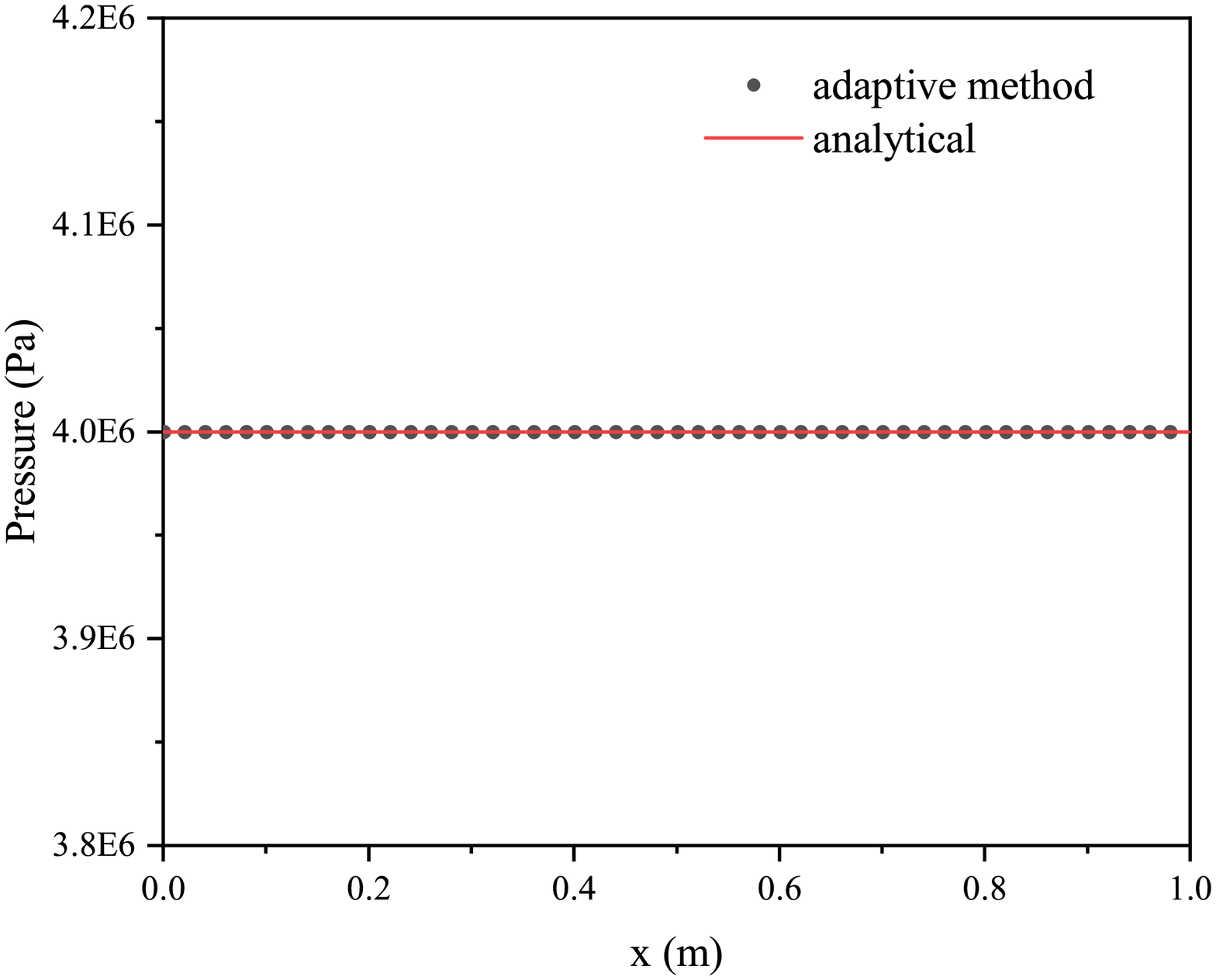}}
	\subfloat[Temperature]{\includegraphics[scale = 0.3]{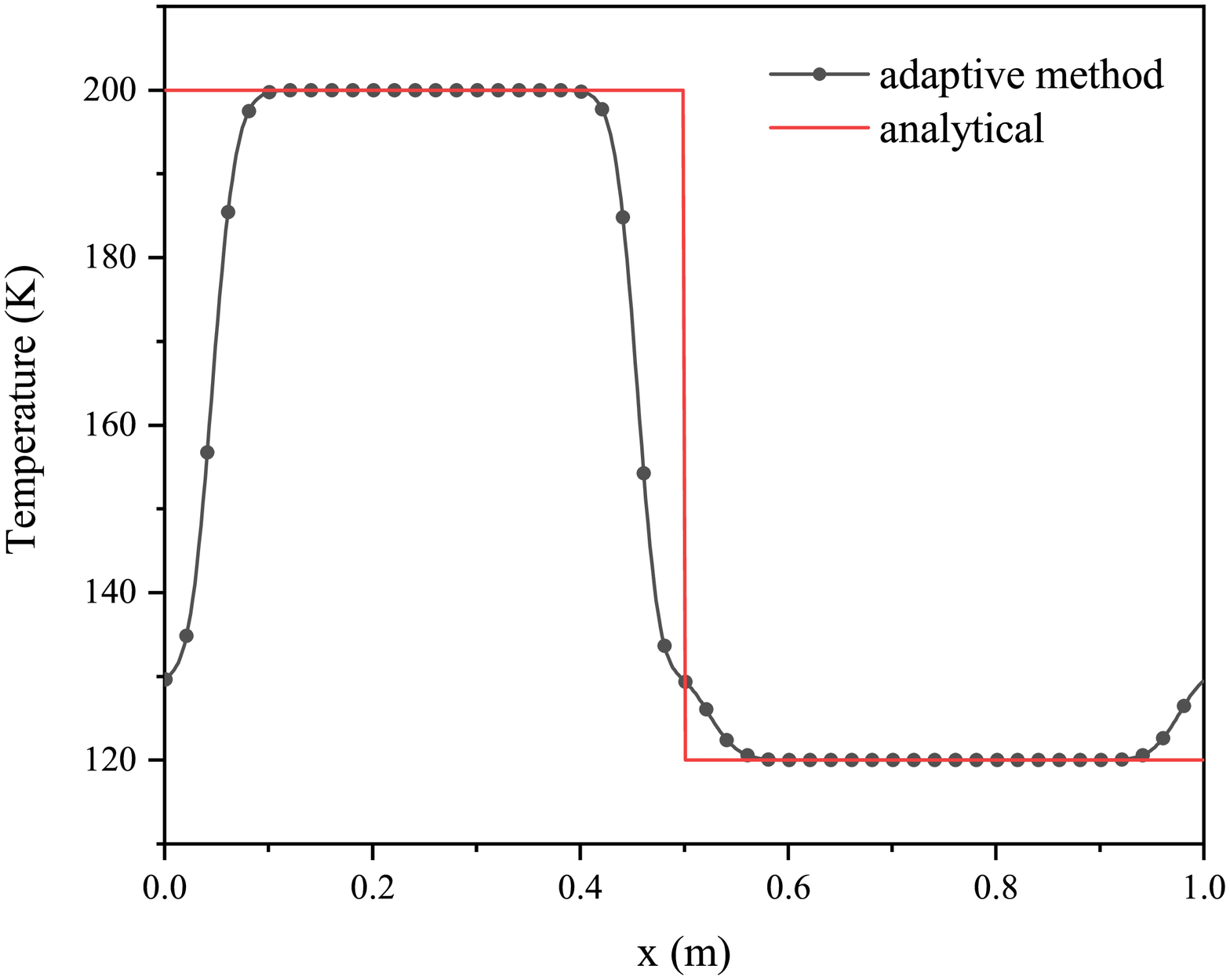}}
	\caption{The profiles of density, velocity, pressure and temperature when resolving 1-D transcritical advection case with sharp interface by adaptive scheme}
	\label{fig::ada}
\end{figure*}

\subsection{Riemann solver for primitive Euler equations}
Most Riemann solvers are developed for the conservative from Euler equations to compute flux at the cell interfaces. The PVRS\cite{toro2013riemann} and Dumbser–Osher–Toro Riemann solver (DOTRS)\cite{dot,ad_osher} can be directly applied in the flux computation for Euler equations in primitive form. The primitive form of DOTRS is used in this paper, in which the tricky path integral is solved numerically by the Gauss–Legendre quadrature rule. The numerical flux at interface is decomposed into a positive part and a negative part:
\begin{align}
		\mathbf{H}_{i+1/2}^{\pm} &= \int_{\mathbf{W}_L}^{\mathbf{W}_R}\mathbf{B}^{\pm}(\mathbf{W}) 
		= \left ( \int_{0}^{1}  \mathbf{B}^{\pm}(\Psi (s; \mathbf{W}_L, \mathbf{W}_R))  ds \right) \Delta \mathbf{W} \notag  \\
		&=\left ( \sum_{j=1}^{N} \omega_j \mathbf{B}^{\pm}(\Psi (s; \mathbf{W}_L, \mathbf{W}_R))  \right) \Delta \mathbf{W}
\end{align}
where $ B^+ $ and $ B^- $ can be written as:
\begin{equation}
	\begin{split}
		\mathbf{B}^{+} &= \frac{1}{2}(\mathbf{B} +|\mathbf{B}|) = (\mathbf{K} \boldsymbol{\Lambda} \mathbf{K}^{-1} + \mathbf{K}\left|  \boldsymbol{\Lambda}\right|  \mathbf{K}^{-1})   \\
		\mathbf{B}^{-} &= \frac{1}{2}(\mathbf{B} +|\mathbf{B}|) = (\mathbf{K} \boldsymbol{\Lambda} \mathbf{K}^{-1} - \mathbf{K}\left|  \boldsymbol{\Lambda}\right|  \mathbf{K}^{-1})
	\end{split}
\end{equation}
Following the paper published by Dumbser and Toro\cite{dot}, the three-point Gauss–Legendre rule is employed for numerical integral:
\begin{equation}
	s_{1,3} = \frac{1}{2} \mp \frac{\sqrt{15}}{10}, \quad s_{2} = \frac{1}{2}, \quad \omega_{1,3} = \frac{5}{18}, \quad \omega_2 = \frac{8}{18}
	\label{eq::gauss}
\end{equation}

Finally, the primitive variables are updated according:
\begin{equation}
	\mathbf{W}_{i}^{n+1} = \mathbf{W}_{i}^{n} - \frac{\Delta t}{\Delta x}(\mathbf{H}_{i+1/2}^{-} + \mathbf{H}_{i-1/2}^{+})
	\label{eq::pvup}
\end{equation}

In the primitive form of DOTRS, only the speed of sound is related to the EoS and the type of EoS is not specified.

\subsection{Riemann solver for real gas}
Riemann solvers for real gas are usually relevant to the partial derivatives of thermophysical properties. However, computing the thermodynamic derivatives in transcritical region using non-linear EoS are often inaccurate and computationally expensive, resulting in unphysical results like negative density. Similar issues arise when using the Roe scheme to real gases. An extended Roe Riemann solver\cite{extension_Roe} developed in recent years is used in this paper to overcome this problem for the simulation of transcritical flows. The numerical flux in Roe Riemann solver can be compute as:
\begin{equation}
	\mathbf{F}_{i+\frac{1}{2}} = \frac{1}{2} \left( \mathbf{F}_{L} + \mathbf{F}_{R}\right)  - \frac{1}{2} \left( \sum_{i=1} ^{m} \tilde{\alpha}_{i} |\tilde{\lambda}_i| \tilde{\mathbf{K}}^{(i)}\right)  
	\label{Roe_flux}
\end{equation}

Here, $ \mathbf{F}_{\mathbf{i}+\frac{\mathbf{1}}{\mathbf{2}}} $ represent fluxes of conservative variables at the interface between cell $ i $ and $ i+1 $. $ \widetilde{\alpha_i}, \widetilde{\lambda_i} $ and $ \widetilde{\mathbf{K}}^{\left(i\right)} $ are wave strengths, eigenvalues, and right eigenvectors of Jacobi matrix computed by Roe average state. The relevant averages of the extended Roe scheme for real gas are given as follows:
\begin{equation}
	\begin{split}
		\tilde{\rho} &= \sqrt{\rho_L \rho_R}   \\
		\tilde{u} &= \frac{\sqrt{\rho_L}u_L + \sqrt{\rho_R}u_R}{\sqrt{\rho_L} + \sqrt{\rho_R}}   \\
		\tilde{H} &= \frac{\sqrt{\rho_L}H_L + \sqrt{\rho_R}H_R}{\sqrt{\rho_L} + \sqrt{\rho_R}} 
	\end{split}
\end{equation}

In order to ensure that the thermodynamic relationship of Roe average state is compatible, the Roe average values of pressure P and temperature T are solved by solving the following equation:
\begin{equation}
	H - \frac{1}{2}u^2 = \frac{\rho e + P }{\rho}
	\label{eq::enthalpy}
\end{equation}
by Newton iteration, which is the definition of enthalpy.

The wave strengths are:
\begin{equation}
	\begin{split}
		\tilde{\alpha}_1 &= \frac{1}{2\tilde{c}^2}[\Delta p - \tilde{\rho}\tilde{c}\Delta u]  \\
		\tilde{\alpha}_2 &= \Delta \rho - \Delta p / \tilde{c}^2   \\
		\tilde{\alpha}_3 &= \frac{1}{2\tilde{c}^2}[\Delta p + \tilde{\rho}\tilde{c}\Delta u]
		\label{eq::wave_strengths}
	\end{split}
\end{equation}

The eigenvalues of Roe average matrix are:
\begin{equation}
	\tilde \lambda_1 = \tilde u - \tilde a, \quad \tilde \lambda_2  = \tilde u , \quad \tilde \lambda_3 = \tilde u + \tilde a
	\label{eq::eigenvalues}
\end{equation}

The corresponding right eigenvectors are as follows:
\begin{equation}
	\begin{split}
		\tilde{\mathbf{K}}^{(1)} &= \begin{bmatrix} 1 \\  \tilde u - \tilde a  \\  \tilde H-\tilde u \tilde a  \end{bmatrix} ; \\ 
		\tilde{\mathbf{K}}^{(2)} &= \begin{bmatrix} 1 \\ \tilde u  \\ \tilde H - \tilde \rho \tilde c^2 \frac{\partial \tilde e}{\partial\tilde p}  \end{bmatrix} ; \\
		\tilde{\mathbf{K}}^{(3)} &= \begin{bmatrix} 1 \\ \tilde u+ \tilde a \\ \tilde H+\tilde u \tilde a  \end{bmatrix}
	\end{split}
\end{equation}

However, when calculating the partial differential of internal energy in the last term of $ \tilde{\mathbf{K}}^{(2)} $ with complicated nonlinear EoS, difficulties always arise. Using the extended Roe scheme for real gas, this term can be solved directly by the definition of Roe scheme: $ F\left(U_R\right)-F\left(U_L\right)=\widetilde{A}\left(U_R-U_L\right) $.

We eventually update the conservative variable in shock-affected cells based on the following equation:
\begin{equation}
	\mathbf{U}_{i}^{n+1} = \mathbf{U}_{i}^{n} + \frac{\Delta t}{\Delta x}\left[ \mathbf{F}_{i-\frac{1}{2}} - \mathbf{F}_{i+\frac{1}{2}}\right] 
\end{equation}

\subsection{Sensor for shock waves}
The primitive method should be locally corrected by a conservative method to correctly resolve the strength and propagation speed of shock waves. In prior research\cite{ad_toro, ad_toro, ad1}, the PVRS was often utilized to detect the existence of shock waves, which provids a switching strategy for the adaptive method.
In PVRS the Riemann problem is approximately solved by\cite{toro2013riemann}:
\begin{equation}
	\begin{split}
		P_{i+\frac{1}{2}} &= \frac{1}{2}\left( p_i + p_{i+1}\right)  + \frac{1}{2}\left( u_i - u_{i+1}\right) \bar{\rho}  \bar{c}   \\
		u_{i+\frac{1}{2}} &= \frac{1}{2}\left( u_i + u_{i+1}\right)  + \frac{1}{2}\left( p_i - p_{i+1}\right) /(\bar{\rho}\bar{c})   \\
		\rho_{i+\frac{1}{2}}^{L} &= \rho_i +\left( u_i - u_{i+\frac{1}{2}}\right) \left( \bar{\rho}/\bar{c}\right)    \\
		\rho_{i+\frac{1}{2}}^{R} &= \rho_{i+1} + \left( u_{i+\frac{1}{2}} - u_{i+1}\right) (\bar{\rho}/\bar{c})
	\end{split}
\end{equation}
with:
\begin{equation}
	\bar{\rho} = \frac{1}{2}(\rho_{i}+\rho_{i+1})  \qquad \bar{c} = \frac{1}{2}(c_i+c_{i+1})
\end{equation}

The propagation speeds of left and right waves are:
\begin{equation}
	\begin{split}
		s_{i+\frac{1}{2}}^{L} &= \frac{\left( \rho_{i}u_{i} - \rho_{i+\frac{1}{2}}^{L}u_{i+\frac{1}{2}}\right) }{\rho_{i} - \rho_{i+\frac{1}{2}}^{L}}    \\
		s_{i+\frac{1}{2}}^{R} &= \frac{\left( \rho_{i+1}u_{i+1} - \rho_{i+\frac{1}{2}}^{R}u_{i+\frac{1}{2}}\right) }{\rho_{i+1} - \rho_{i+\frac{1}{2}}^{R}}
	\end{split}
\end{equation}

Similar to previous research\cite{ad_toro, ad_toro, ad1,toro2013riemann}, the conservative methods are applied at cell $ i $ when:
\begin{equation}
	\begin{split}
		\frac{P_{i+\frac{1}{2}}}{P_i} >& 1+\varepsilon \quad \text{and} \quad s_{i+\frac{1}{2}}^{L} <0  \\
		\text{or} \quad \frac{P_{i-\frac{1}{2}}}{P_i} >& 1+\varepsilon \quad \text{and} \quad s_{i-\frac{1}{2}}^{R} <0
	\end{split}
\end{equation}
where the parameter $ \varepsilon $ can be selected within the range $  (0, 0.1) $, and in this paper $ \varepsilon=0.05 $ is adopted.
\begin{figure}[h]
	{\includegraphics[scale=0.3]{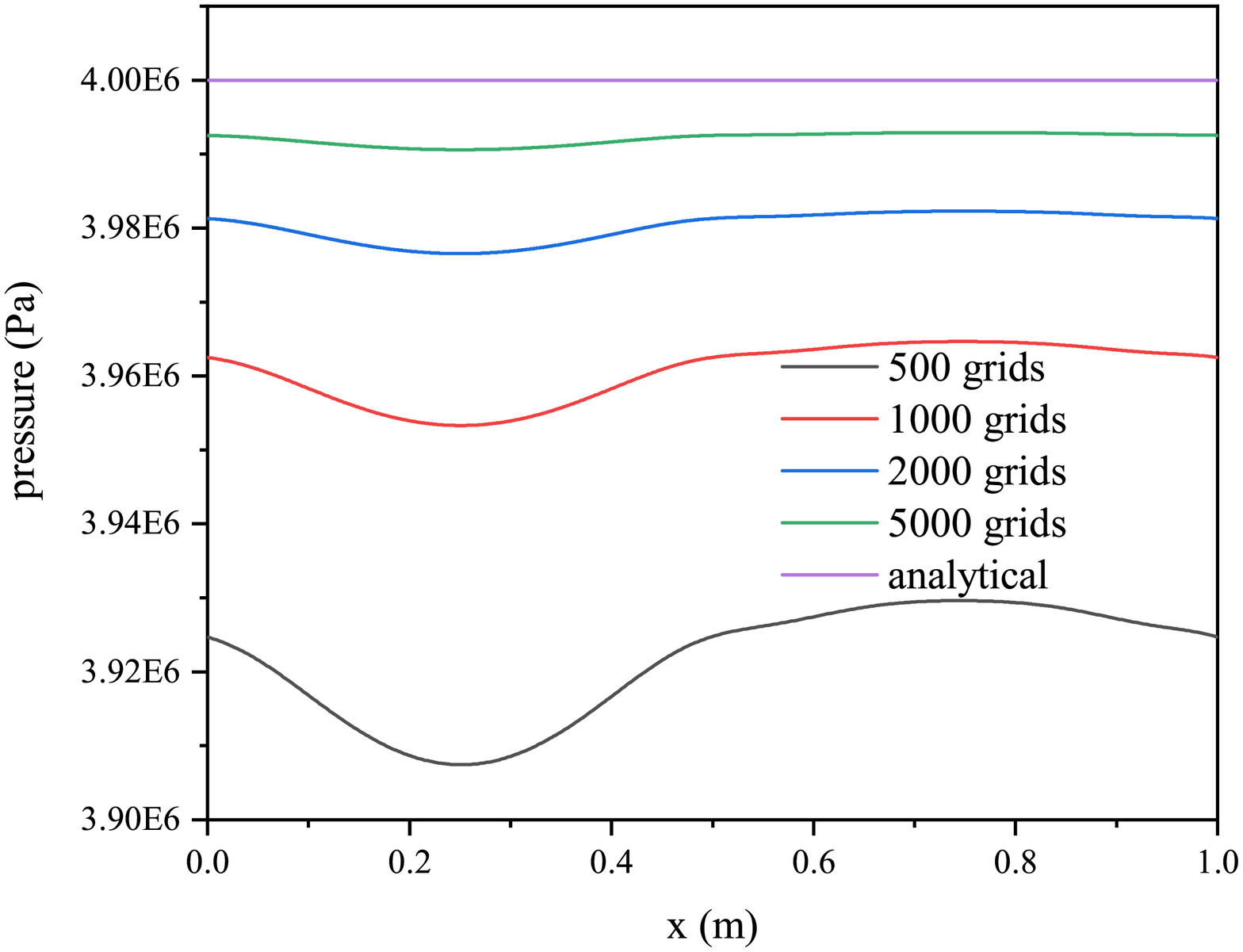}}
	\caption{Results of pressure when solving transcritical advection case with smooth initial condition by conservative scheme}
	\label{fig:con_smooth}
\end{figure}
\begin{figure}[h]
	{\includegraphics[scale=0.3]{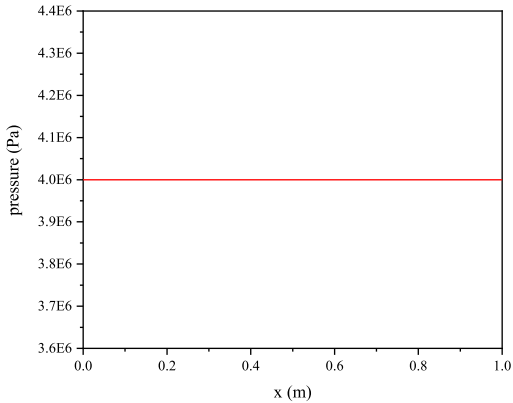}}
	\caption{Results of pressure when solving transcritical advection case with smooth initial condition by primitive scheme}
	\label{fig:ad_smooth}
\end{figure}
\section{Numerical test}
In this section, a series of 1-D and 2-D numerical tests are undertaken to illustrate the capacity of the proposed method in the simulation of transcritical flow. The transcritical nitrogen is adopted as working fluid in all of the test cases with the thermodynamic properties of fluids obtained by PR EoS. The numerical results are compared with available standard solution. For the purpose of analyzing the scheme without the influence of high-order reconstructions, all the results are computed using first order upwind Godunov method, and the extension to high order scheme is straightforward. The following cases demonstrate that the hybrid method suggested in this paper can obtain both oscillation free results as well as correct shock wave strength in transcritical region.

\begin{figure*}[h!t]
	\centering
	\subfloat[Density]{\includegraphics[scale=0.3]{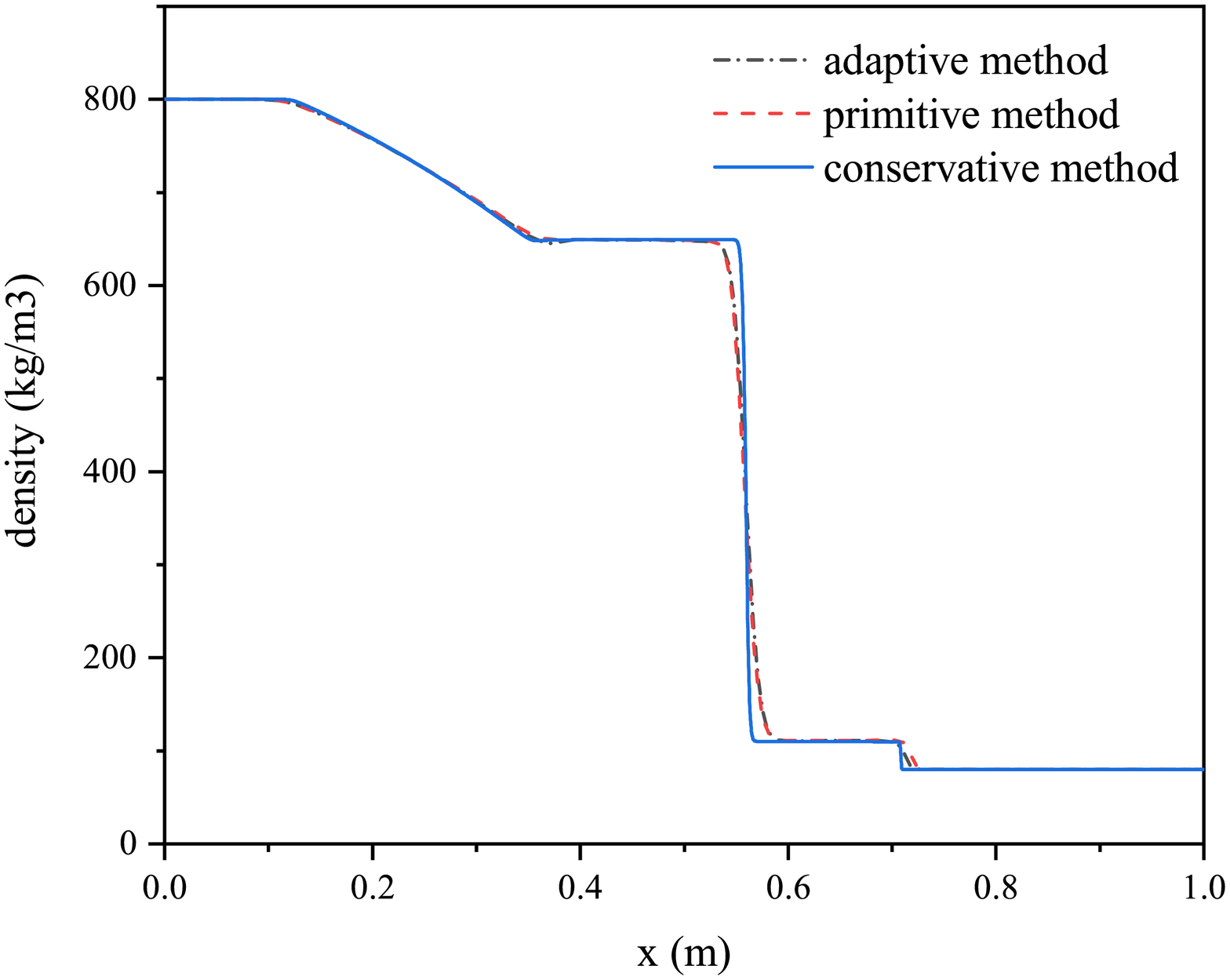}}
	\subfloat[Velocity]{\includegraphics[scale=0.3]{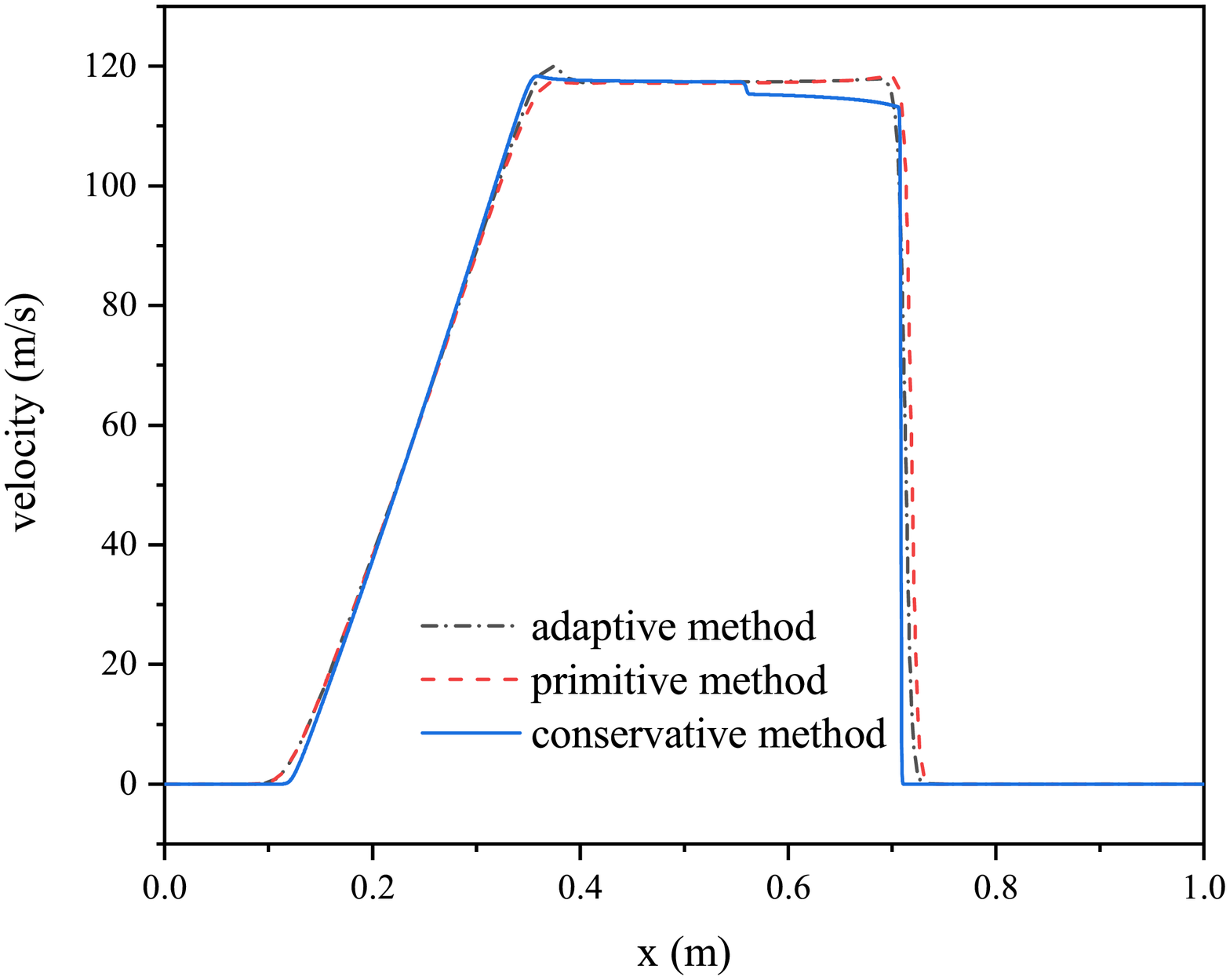}}
	\quad
	\subfloat[Pressure]{\includegraphics[scale = 0.3]{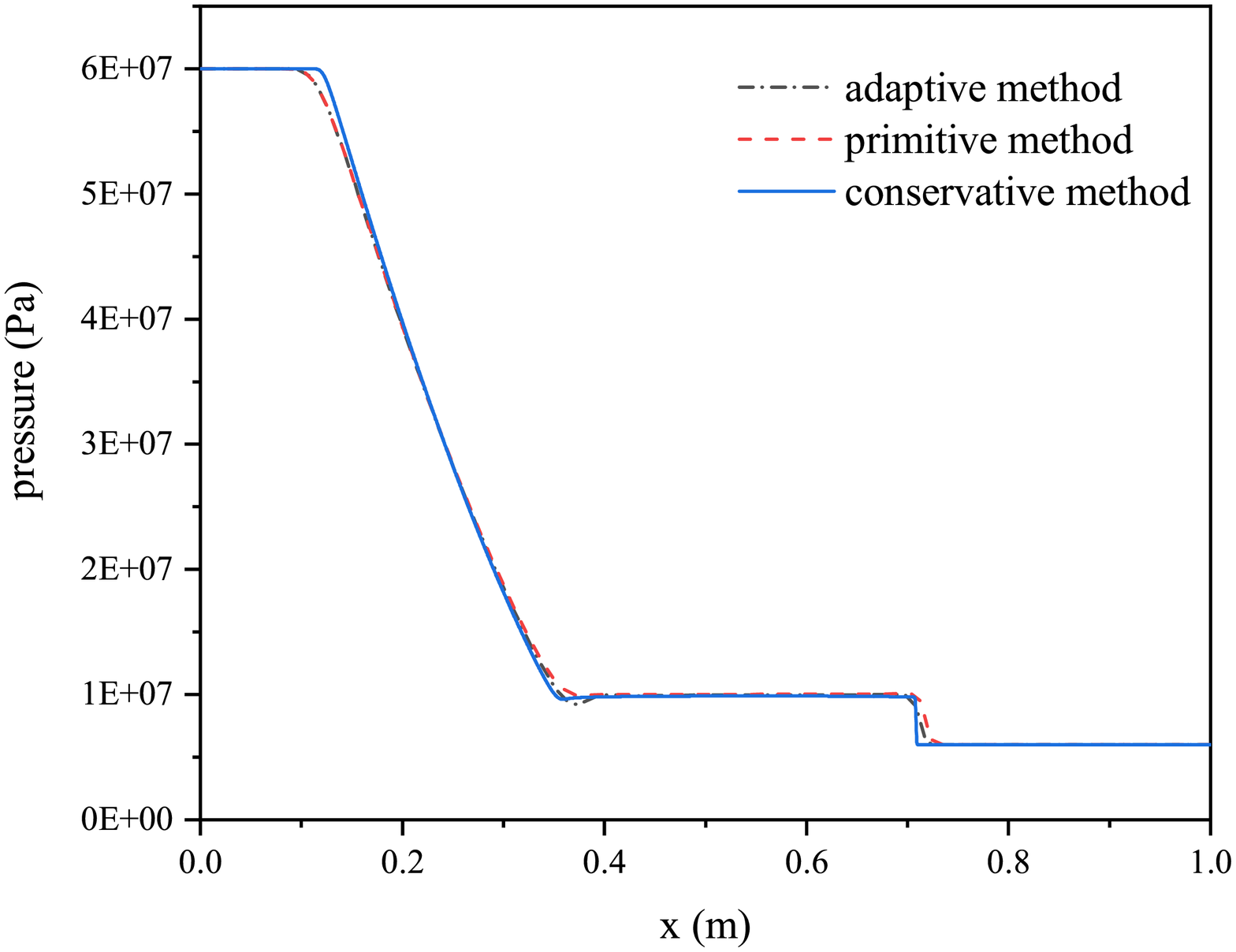}}
	\subfloat[Temperature]{\includegraphics[scale = 0.3]{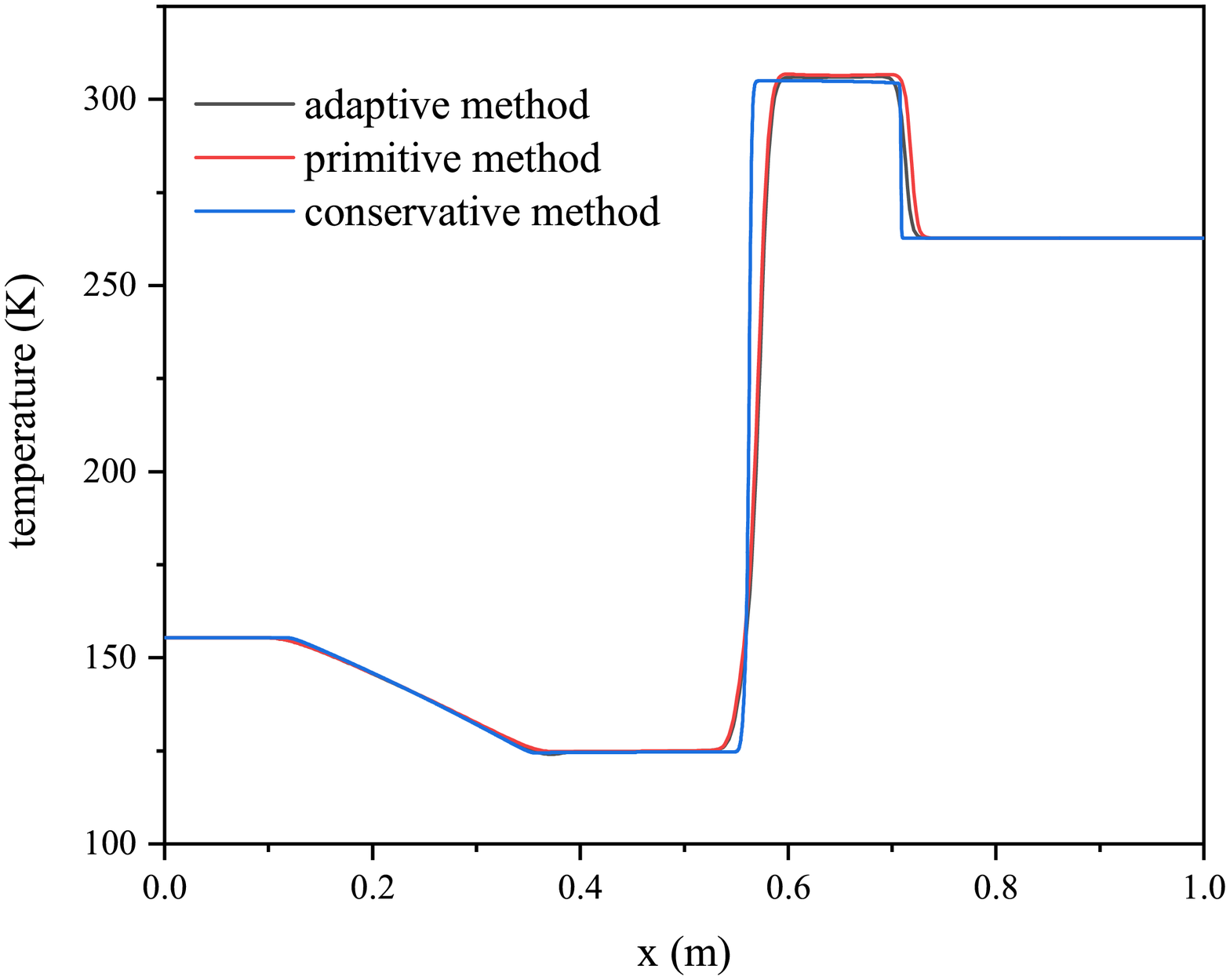}}
	\caption{Profiles of density, velocity, pressure and temperature for the transcritical shock tube problems resolved by different schemes}
	\label{fig::shocktube}
\end{figure*}

\subsection{1-D transcritical advection cases}
In the advection cases, the discontinuous initial condition, which includes sharp transcritical interface, and smooth initial condition are taken into consideration. Compared to conservative method, the hybrid method can produce oscillation free results under any initial conditions. The computational domain is$  x\in\left[0,1\right] $, and both ends have periodic boundary conditions. The CFL number is $ 0.8 $.

The initial values of density, velocity pressure, and temperature at the left and right sides of the interface in the transcritical advection case with discontinuous initial condition are:
\begin{equation}
	\begin{split}
	\begin{bmatrix} \rho_l \\ u_l \\ p_l \\ T_l \end{bmatrix} &= \begin{bmatrix} 580.586\ kg/ m^3 \\ 100\ m/s \\ 4e6\ Pa \\ 120\ K\end{bmatrix} \quad \text{if } 0 \le x \le 0.5  \\
	\begin{bmatrix} \rho_r \\ u_r \\ p_r \\ T_r\end{bmatrix} &= \begin{bmatrix} 74.7415\ kg/ m^3 \\ 100\ m/s \\ 4e6\ Pa \\ 200\ K\end{bmatrix} \quad \text{if } 0.5 < x \le 1.0
	\label{eq::initial_sharp}
	\end{split}
\end{equation}

The sharp interface is designed with a liquid-like state on one side and a gas-like state on the other. At the transcritical interface, there is a discontinuity in temperature and density, but pressure and velocity remain constant across the computational domain. In fig.\ref{fig::adcon} and Fig.[], we present the numerical results of density, velocity, pressure and temperature for the above advection problem resolved using conservative and adaptive methods respectively. Large oscillations can be observed when using conservative method, resulting in the failure of this frequently employed method. The underlying reason for these oscillations is that the pressure variations arise around the interface due to the nonlinearity of thermodynamics, subsequently generating waves and destroying the flow field. There is no oscillation in the numerical results while using the adaptive method, indicating the robustness of the suggested method in the context of transcritical flow. The non-smoothness of the temperature curve is a result of the nonlinear relationship between density and temperature in the transcritical region, and not the impact of the numerical scheme.

In the advection case with smooth initial condition, the values of density, velocity, and pressure are given by:
\begin{equation}
	\begin{split}
	\rho &= 0.5*(\rho_{l}+\rho_{r}) + 0.5*(\rho_r - \rho_l) * \sin(2\pi x) \\
	u &= 100\ m/s \\
	p &= 4e6\ Pa
	\label{eq::initial_smooth}
	\end{split}
\end{equation}
The values of $ \rho_l $ and $ \rho_r $ in $\eqref{eq::initial_smooth}$ are same to $\eqref{eq::initial_sharp}$.In fig.\ref{fig:con_smooth}, we illustrate the numerical results of pressure profile simulated by conservative schemes with different mesh numbers. The pressure profile resolved by hybrid scheme is shown in fig.\ref{fig:ad_smooth}. Similar to the situation in sharp interface case, hybrid scheme can produce numerical results without oscillations. When the initial conditions are smooth, the oscillations in the results resolved by conservative method steadily decrease as the number of grids increases but the oscillations only alleviate never vanish. This indicates that when the initial conditions are smooth, the numerical oscillations of the conservative scheme can be alleviated by refining the mesh so that the relationship between the thermodynamic parameters within a cell can be linearly approximated. However, it seems to be difficult in application for two reasons. First, in recent experimental studies\cite{internal1} has shown that that the transcritical layer is often just a few tens of microns thick. Second, the contact discontinuity in supersonic flow is a sharp interface with temperature and density discontinuities. The hybrid scheme provides a possible way for the robust simulation in transcritical flow with large gradient or discontinuities.

\begin{figure}[h!]
	{\includegraphics[scale=0.3]{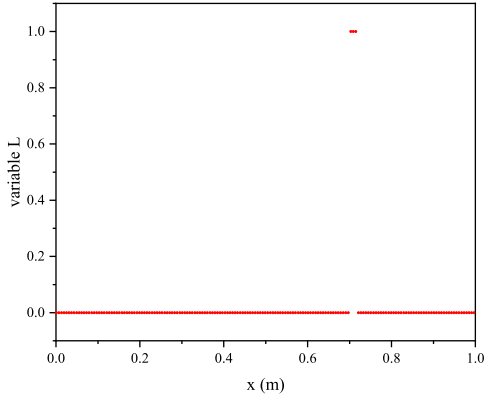}}
	\caption{Variable $ L $}
	\label{fig:shockcp}
\end{figure}
\begin{figure}[h!]
	{\includegraphics[width=0.9\linewidth]{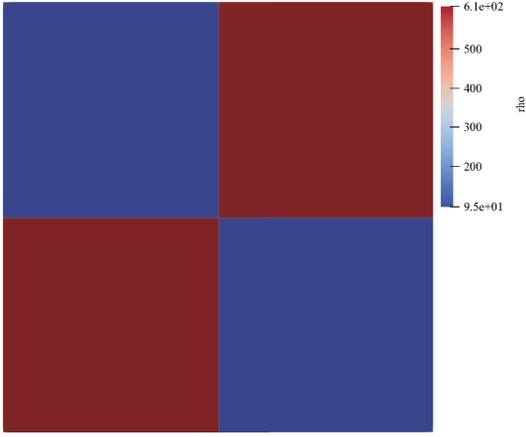}}
	\caption{Initial conditions of density for the 2-D transcritical advection test case}
	\label{fig:init_2Dad}
\end{figure}

\subsection{1-D shock tube problems}
In this section, we employ the 1-D shock tube problems to demonstrate that the hybrid scheme is capable of accurately resolving the strength and propagation speed of shock waves. The computational domain is $ x\in\left[0,1 \ m\right] $, with transmissive boundary conditions in both ends. The CFL number is set to $ 0.8 $, and the simulation time is $ t = 5e-4s $.  The initial conditions for 1-D transcritical shock tube test case are:
\begin{equation}
	\begin{split}
		\begin{bmatrix} \rho_l \\ u_l \\ p_l\end{bmatrix} &= \begin{bmatrix} 800\ kg/m^3 \\ 0\ m/s \\ 60e6\ Pa\end{bmatrix},  \\
		\begin{bmatrix} \rho_r \\ u_r \\ p_r\end{bmatrix} &= \begin{bmatrix} 80\ kg/m^3 \\ 0\ m/s \\ 6e6\ Pa\end{bmatrix}
	\end{split}
\end{equation}

 Fig.\ref{fig::shocktube} illustrates the density, velocity, pressure, and temperature profiles for the transcritical shock tube problem resolved using the adaptive, primitive and conservative schemes. To provide reference locations of discontinuities, the results resolved by conservative scheme are simulated using a mesh with 5000 grids, while the mesh with 500 grids are used for adaptive and primitive schemes. It is evident that the primitive scheme predicts a faster speed of shock waves. The location of shock waves resolved by hybrid method are in good agreement with those predicted by conservative scheme. However, an unphysical discontinuity of velocity can be observed in the location of contact discontinuity when using conservative scheme. The above one-dimension numerical tests demonstrate the robustness and precision of the adaptive primitive-conservative scheme in the simulations of high speed transcritical flow. In Fig.\ref{fig:shockcp}, we present the existence of shock waves predicted by PVRS shock sensor. If the cells are affected by shock waves, the variable $ L $ will equal to one. One can clearly observed that shock waves are correctly predicted by the PVRS shock sensor. Thus, conservative method will be used in the shock involved cells, which maintain the adaptive scheme correctly resolving the propagation speed and strength of shock waves.

\begin{figure}[h!t]
	\centering
	\subfloat[Density]{\includegraphics[width=0.9\linewidth]{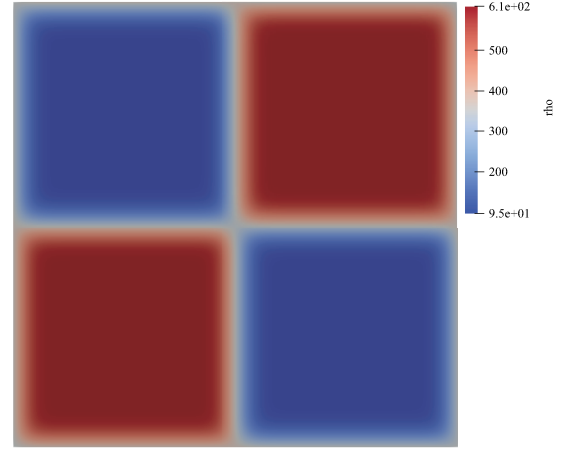}}
	\quad
	\subfloat[Pressure]{\includegraphics[width=0.9\linewidth]{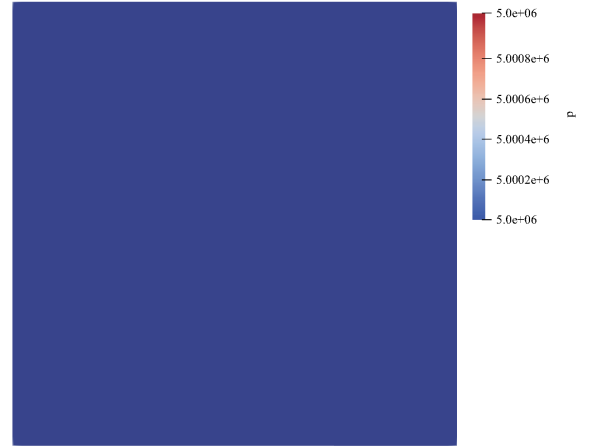}}
	\caption{Distribution of density and pressure for the 2-D transcritical advection test case at time $ t = 5e-3s $}
	\label{fig::2Dad}
\end{figure}

\begin{figure*}[h!t]
	\centering
	\subfloat[Density at time $t = 4.46e-4s$]{\includegraphics[width=0.45\linewidth]{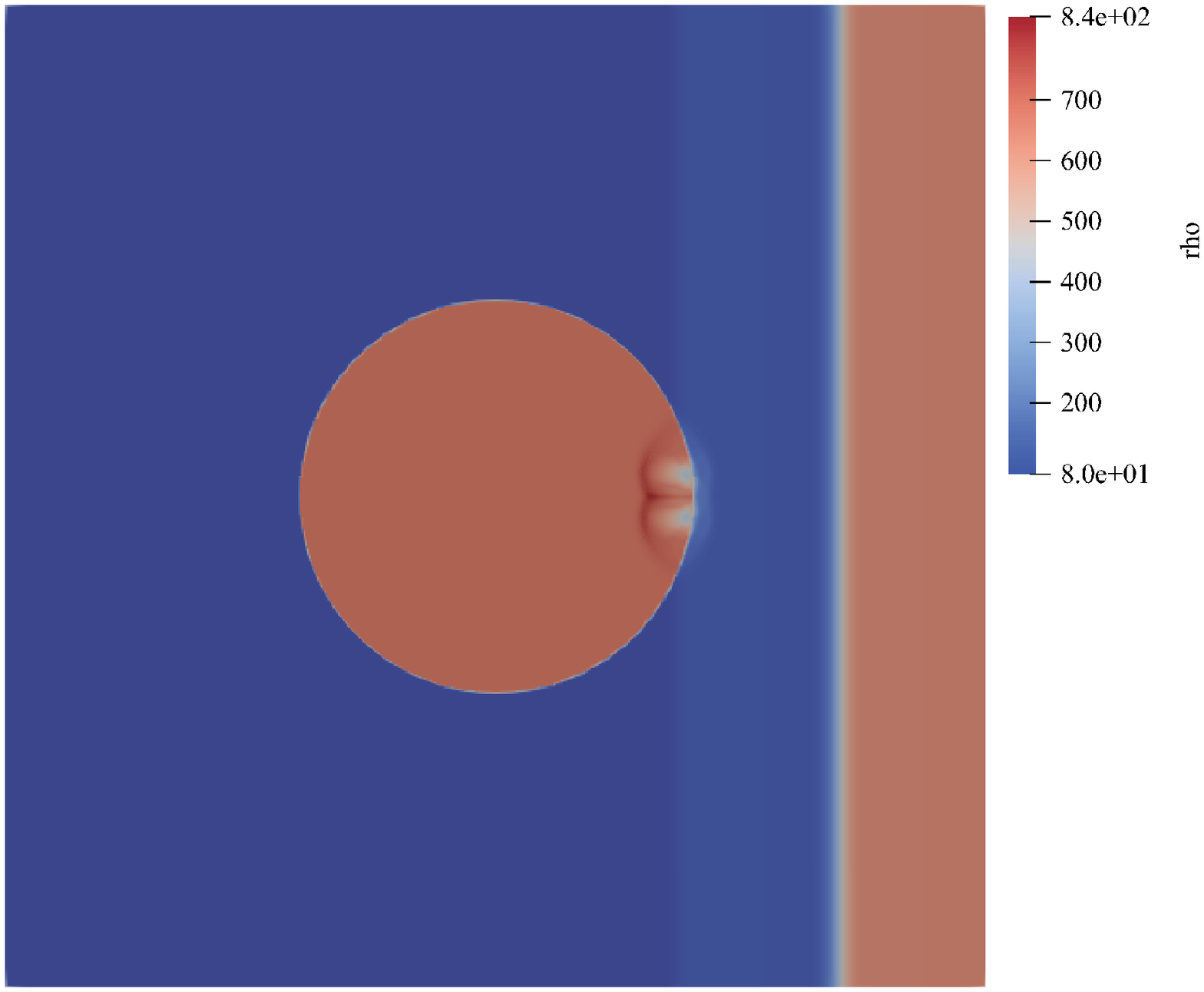}}
	\subfloat[Pressure at time $t = 4.46e-4s$]{\includegraphics[width=0.45\linewidth]{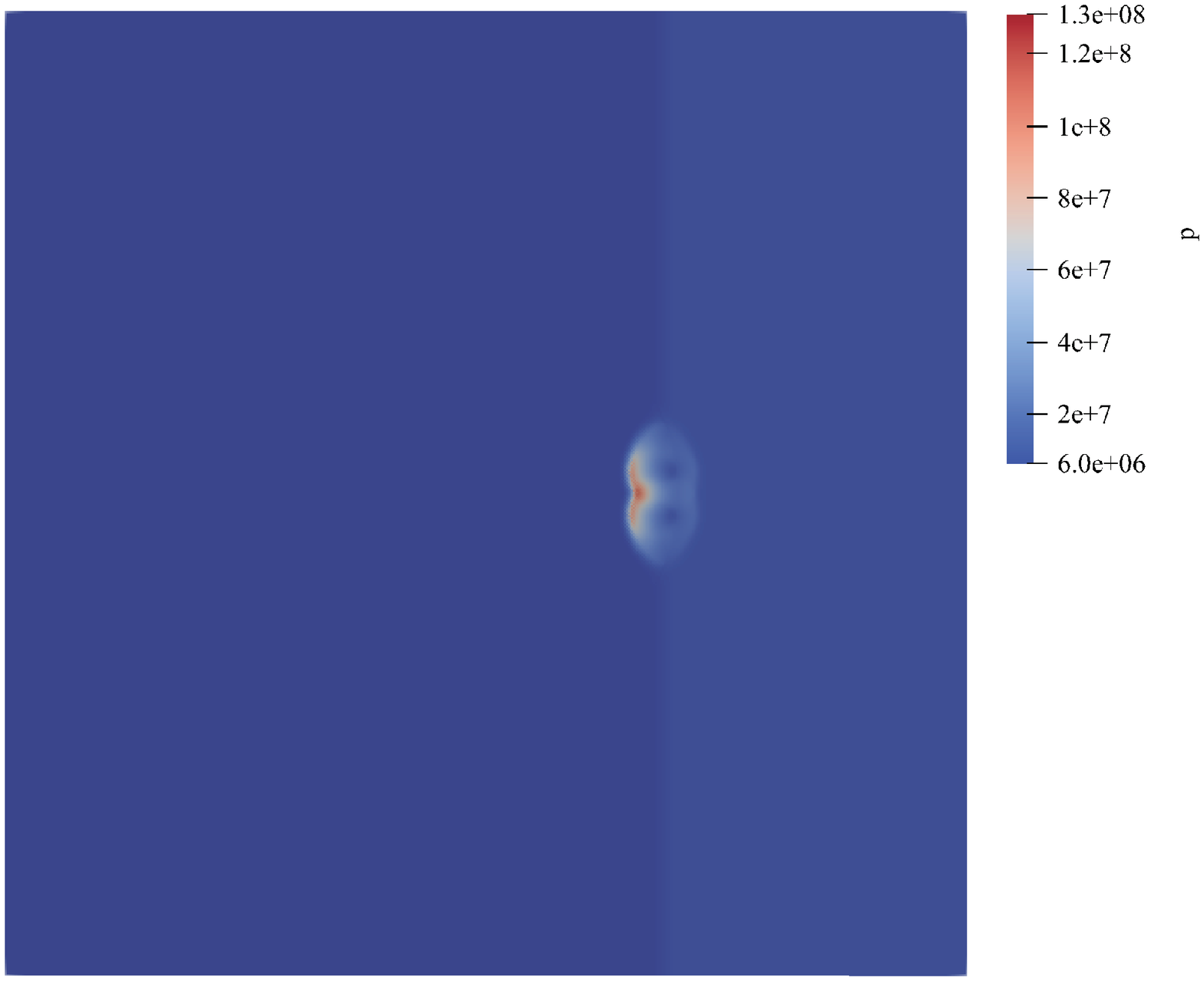}}
	\quad
	\subfloat[Density at time $t = 6.26e-4s$]{\includegraphics[width=0.45\linewidth]{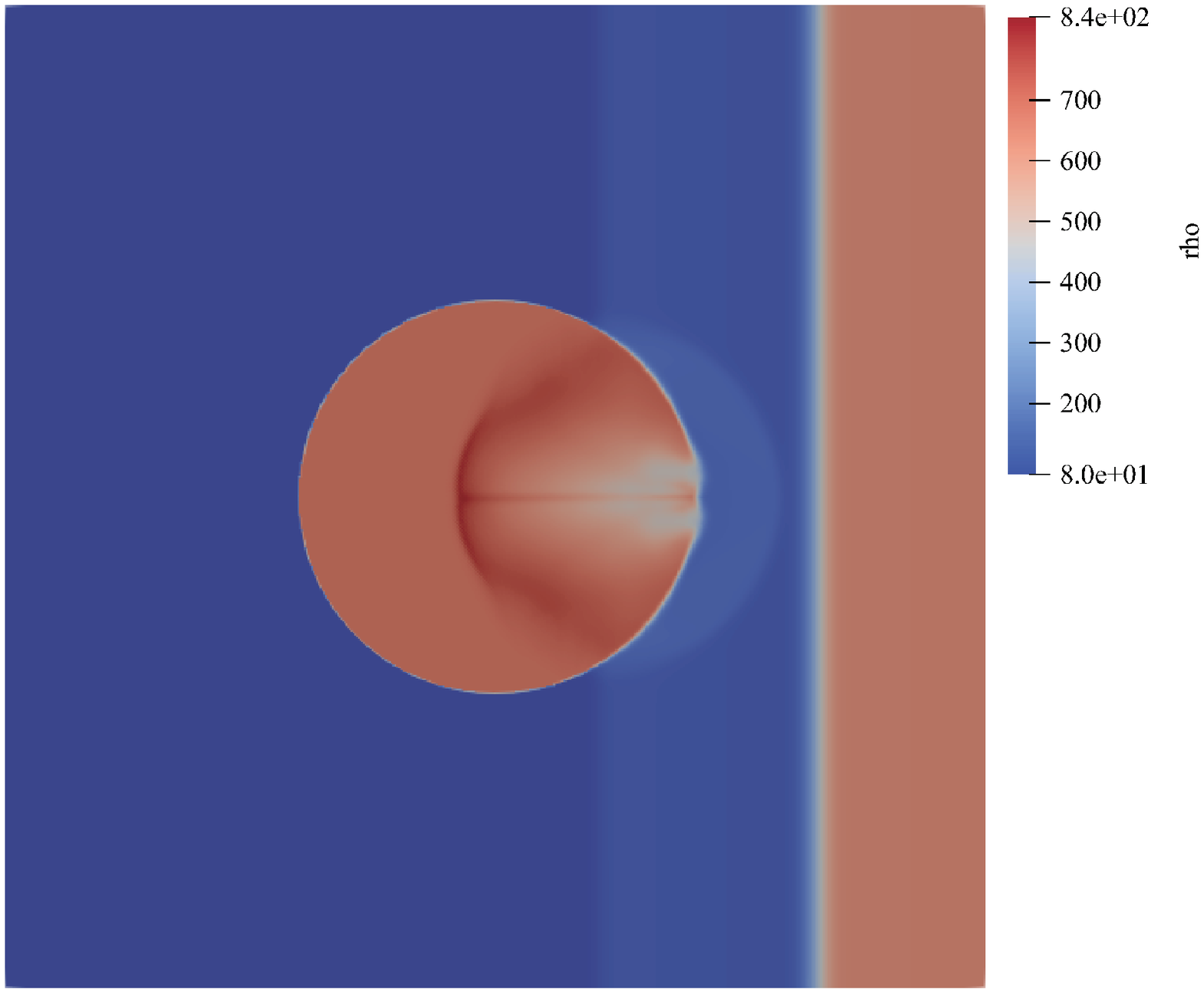}}
	\subfloat[Pressure at time $t = 6.26e-4s$]{\includegraphics[width=0.45\linewidth]{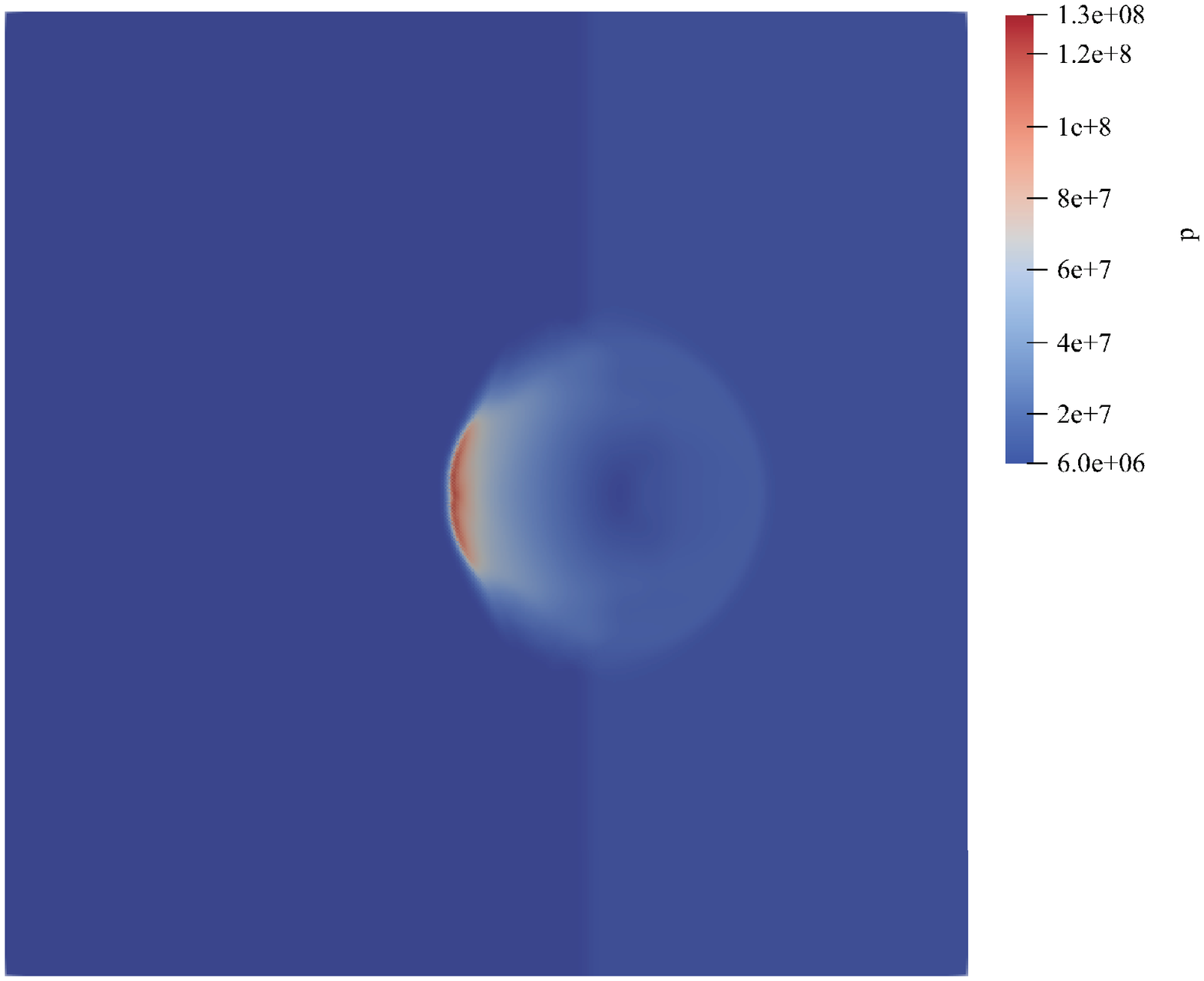}}
	\caption{Density and pressure distributions of shock and transcritical droplet interaction problem at different time}
	\label{fig::shockbubble}
\end{figure*}

\subsection{2-D transcritical advection test case}
In this section, the hybrid scheme is applied to a 2-D advection test case. The computational domain is taken$  [0, 1m] * [0, 1m] $ with periodic boundary condition in both directions. The initial pressure $ P=5e6Pa $ and velocity $ [u, v] = [50m/s, 50m/s] $ (diagonal flow) are uniform in the whole computational domain. The CFL number is $ 0.5 $. For the purpose of illustrating the effect numerical schemes itself, first order upwind Godunov scheme are used in the simulation. In Fig.\ref{fig:init_2Dad}, we illustrate the initial condition of density and the sketch of computational domain. The exact solution of Euler equations with above condition is clearly just a passive convection of density and temperature. Fig.\ref{fig::2Dad} shows the distribution of density and pressure at time $ t = 5e-3s $. Similar to 1-D convection case, oscillation free results are resolved by the hybrid scheme in 2-D transcritical convection numerical test. The above results indicate that the adaptive method is capable of getting a robust and accurate result in the simulation of transcritical flows.

\subsection{Interaction between shock waves and transcritical droplet}
In this section, the adaptive scheme is employed to resolve the interaction between shock waves and transcritical droplet. The size of computational domain is set to $ [0, 1m] * [0, 1m] $, with transmissive boundary conditions in both ends. A transcritical droplet with a radius of $ r = 0.2m $ locates in the center of computational domain. The initial velocities in x and y directions are set to zero over the whole computational domain. The initial conditions for density, velocity and pressure in the different region of computational domain are as follows:
\begin{equation}
	\begin{split}
		\begin{bmatrix} \rho_{drop} \\ u_{drop} \\ v_{drop} \\ P_{drop} \end{bmatrix} &= \begin{bmatrix} 700\ kg/ m^3 \\ 0\ m/s \\ 0\ m/s \\ 6e6\ Pa\end{bmatrix}   \\
		\begin{bmatrix} \rho_{left} \\ u_{left} \\ v_{left} \\ P_{left} \end{bmatrix} &= \begin{bmatrix} 700\ kg/ m^3 \\ 0\ m/s \\ 0\ m/s \\ 6e6\ Pa\end{bmatrix}   \\
		\begin{bmatrix} \rho_{right} \\ u_{right} \\ v_{right} \\ P_{right} \end{bmatrix} &= \begin{bmatrix} 800\ kg/ m^3 \\ 0\ m/s \\ 0\ m/s \\ 60e6\ Pa\end{bmatrix}
	\end{split}
\end{equation}

Initially a discontinuity is positioned at x = 0.9, with pressure on the right side being greater than that on the left side, producing a shock wave to propagate to the left. Fig[12] depicts the pressure contours at various times. No oscillation can be observed near the boundary of transcritical droplet, which means that the robust result can be obtained by the adaptive method.

\section{Conclusion }
We have developed an adaptive primitive-conservative scheme for supersonic transcritical flows with an arbitrary equation of state without calculating the derivatives of thermodynamic values. One- and two-dimensional numerical tests are employed to illustrate the accuracy and robustness of the adaptive scheme. Even when transcritical discontinuities and shock waves coexist, the adaptive strategy seems to work well. In regions with high nonlinear thermodynamics, the adaptive scheme eliminates pressure oscillations. In addition, the strength and propagation speed of shock waves are accurately determined by locally adapting the conservative scheme in the cells impacted by the shock waves.

\section*{Acknowledgements}
The authors gratefully acknowledge the financial support of the Natural Science Fund of China (Grant No: 91741103) and the National Key Research and Development Program of China (Grant No: 2016YFB0600101).

\section*{AUTHOR DECLARATIONS}
\subsection*{Conflict of Interest}
The authors have no conflicts to disclose.
\subsection*{DATA AVAILABILITY}
The data that support the findings of this study are available from the corresponding author upon reasonable request.

\section*{REFERENCES}
\bibliography{aipsamp}

\begin{thebibliography}{42}%
\makeatletter
\providecommand \@ifxundefined [1]{%
 \@ifx{#1\undefined}
}%
\providecommand \@ifnum [1]{%
 \ifnum #1\expandafter \@firstoftwo
 \else \expandafter \@secondoftwo
 \fi
}%
\providecommand \@ifx [1]{%
 \ifx #1\expandafter \@firstoftwo
 \else \expandafter \@secondoftwo
 \fi
}%
\providecommand \natexlab [1]{#1}%
\providecommand \enquote  [1]{``#1''}%
\providecommand \bibnamefont  [1]{#1}%
\providecommand \bibfnamefont [1]{#1}%
\providecommand \citenamefont [1]{#1}%
\providecommand \href@noop [0]{\@secondoftwo}%
\providecommand \href [0]{\begingroup \@sanitize@url \@href}%
\providecommand \@href[1]{\@@startlink{#1}\@@href}%
\providecommand \@@href[1]{\endgroup#1\@@endlink}%
\providecommand \@sanitize@url [0]{\catcode `\\12\catcode `\$12\catcode
  `\&12\catcode `\#12\catcode `\^12\catcode `\_12\catcode `\%12\relax}%
\providecommand \@@startlink[1]{}%
\providecommand \@@endlink[0]{}%
\providecommand \url  [0]{\begingroup\@sanitize@url \@url }%
\providecommand \@url [1]{\endgroup\@href {#1}{\urlprefix }}%
\providecommand \urlprefix  [0]{URL }%
\providecommand \Eprint [0]{\href }%
\providecommand \doibase [0]{http://dx.doi.org/}%
\providecommand \selectlanguage [0]{\@gobble}%
\providecommand \bibinfo  [0]{\@secondoftwo}%
\providecommand \bibfield  [0]{\@secondoftwo}%
\providecommand \translation [1]{[#1]}%
\providecommand \BibitemOpen [0]{}%
\providecommand \bibitemStop [0]{}%
\providecommand \bibitemNoStop [0]{.\EOS\space}%
\providecommand \EOS [0]{\spacefactor3000\relax}%
\providecommand \BibitemShut  [1]{\csname bibitem#1\endcsname}%
\let\auto@bib@innerbib\@empty
\bibitem [{\citenamefont {Manin}\ \emph {et~al.}(2014)\citenamefont {Manin},
  \citenamefont {Bardi}, \citenamefont {Pickett}, \citenamefont {Dahms},\ and\
  \citenamefont {Oefelein}}]{internal1}%
  \BibitemOpen
  \bibfield  {author} {\bibinfo {author} {\bibfnamefont {J.}~\bibnamefont
  {Manin}}, \bibinfo {author} {\bibfnamefont {M.}~\bibnamefont {Bardi}},
  \bibinfo {author} {\bibfnamefont {L.}~\bibnamefont {Pickett}}, \bibinfo
  {author} {\bibfnamefont {R.}~\bibnamefont {Dahms}}, \ and\ \bibinfo {author}
  {\bibfnamefont {J.}~\bibnamefont {Oefelein}},\ }\bibfield  {title} {\enquote
  {\bibinfo {title} {Microscopic investigation of the atomization and mixing
  processes of diesel sprays injected into high pressure and temperature
  environments},}\ }\href {\doibase https://doi.org/10.1016/j.fuel.2014.05.060}
  {\bibfield  {journal} {\bibinfo  {journal} {Fuel}\ }\textbf {\bibinfo
  {volume} {134}},\ \bibinfo {pages} {531--543} (\bibinfo {year}
  {2014})}\BibitemShut {NoStop}%
\bibitem [{\citenamefont {Genzale}\ \emph {et~al.}(2011)\citenamefont
  {Genzale}, \citenamefont {Idicheria}, \citenamefont {Manin}, \citenamefont
  {Pickett}, \citenamefont {Siebers},\ and\ \citenamefont
  {Musculus}}]{internal2}%
  \BibitemOpen
  \bibfield  {author} {\bibinfo {author} {\bibfnamefont {C.~L.}\ \bibnamefont
  {Genzale}}, \bibinfo {author} {\bibfnamefont {C.~A.}\ \bibnamefont
  {Idicheria}}, \bibinfo {author} {\bibfnamefont {J.}~\bibnamefont {Manin}},
  \bibinfo {author} {\bibfnamefont {L.~M.}\ \bibnamefont {Pickett}}, \bibinfo
  {author} {\bibfnamefont {D.~L.}\ \bibnamefont {Siebers}}, \ and\ \bibinfo
  {author} {\bibfnamefont {M.~P.~B.}\ \bibnamefont {Musculus}},\ }\bibfield
  {title} {\enquote {\bibinfo {title} {Relationship between diesel fuel spray
  vapor penetration/dispersion and local fuel mixture fraction},}\ }\href
  {\doibase https://doi.org/10.4271/2011-01-0686} {\bibfield  {journal}
  {\bibinfo  {journal} {SAE International Journal of Engines}\ }\textbf
  {\bibinfo {volume} {4}},\ \bibinfo {pages} {764--799} (\bibinfo {year}
  {2011})}\BibitemShut {NoStop}%
\bibitem [{\citenamefont {Rodriguez}\ \emph {et~al.}(2019)\citenamefont
  {Rodriguez}, \citenamefont {Rokni}, \citenamefont {Koukouvinis},
  \citenamefont {Gupta},\ and\ \citenamefont {Gavaises}}]{internal3}%
  \BibitemOpen
  \bibfield  {author} {\bibinfo {author} {\bibfnamefont {C.}~\bibnamefont
  {Rodriguez}}, \bibinfo {author} {\bibfnamefont {H.~B.}\ \bibnamefont
  {Rokni}}, \bibinfo {author} {\bibfnamefont {P.}~\bibnamefont {Koukouvinis}},
  \bibinfo {author} {\bibfnamefont {A.}~\bibnamefont {Gupta}}, \ and\ \bibinfo
  {author} {\bibfnamefont {M.}~\bibnamefont {Gavaises}},\ }\bibfield  {title}
  {\enquote {\bibinfo {title} {Complex multicomponent real-fluid thermodynamic
  model for high-pressure diesel fuel injection},}\ }\href {\doibase
  https://doi.org/10.1016/j.fuel.2019.115888} {\bibfield  {journal} {\bibinfo
  {journal} {Fuel}\ }\textbf {\bibinfo {volume} {257}},\ \bibinfo {pages}
  {115888} (\bibinfo {year} {2019})}\BibitemShut {NoStop}%
\bibitem [{\citenamefont {Candel}, \citenamefont {Schmitt},\ and\ \citenamefont
  {Darabiha}(2011)}]{turbine1}%
  \BibitemOpen
  \bibfield  {author} {\bibinfo {author} {\bibfnamefont {S.}~\bibnamefont
  {Candel}}, \bibinfo {author} {\bibfnamefont {T.}~\bibnamefont {Schmitt}}, \
  and\ \bibinfo {author} {\bibfnamefont {N.}~\bibnamefont {Darabiha}},\
  }\bibfield  {title} {\enquote {\bibinfo {title} {Progress in transcritical
  combustion: experimentation, modeling and simulation},}\ }\href@noop {}
  {\bibfield  {journal} {\bibinfo  {journal} {23rd ICDERS, Irvine}\ }\textbf
  {\bibinfo {volume} {111}} (\bibinfo {year} {2011})}\BibitemShut {NoStop}%
\bibitem [{\citenamefont {Pons}\ \emph {et~al.}(2009)\citenamefont {Pons},
  \citenamefont {Darabiha}, \citenamefont {Candel}, \citenamefont {Ribert},\
  and\ \citenamefont {Yang}}]{turbine2}%
  \BibitemOpen
  \bibfield  {author} {\bibinfo {author} {\bibfnamefont {L.}~\bibnamefont
  {Pons}}, \bibinfo {author} {\bibfnamefont {N.}~\bibnamefont {Darabiha}},
  \bibinfo {author} {\bibfnamefont {S.}~\bibnamefont {Candel}}, \bibinfo
  {author} {\bibfnamefont {G.}~\bibnamefont {Ribert}}, \ and\ \bibinfo {author}
  {\bibfnamefont {V.}~\bibnamefont {Yang}},\ }\bibfield  {title} {\enquote
  {\bibinfo {title} {Mass transfer and combustion in transcritical non-premixed
  counterflows},}\ }\href {\doibase 10.1080/13647830802368821} {\bibfield
  {journal} {\bibinfo  {journal} {Combustion Theory and Modelling}\ }\textbf
  {\bibinfo {volume} {13}},\ \bibinfo {pages} {57--81} (\bibinfo {year}
  {2009})},\ \Eprint
  {http://arxiv.org/abs/https://doi.org/10.1080/13647830802368821}
  {https://doi.org/10.1080/13647830802368821} \BibitemShut {NoStop}%
\bibitem [{\citenamefont {Lefebvre}\ and\ \citenamefont
  {Ballal}(2010)}]{turbine3}%
  \BibitemOpen
  \bibfield  {author} {\bibinfo {author} {\bibfnamefont {A.~H.}\ \bibnamefont
  {Lefebvre}}\ and\ \bibinfo {author} {\bibfnamefont {D.~R.}\ \bibnamefont
  {Ballal}},\ }\href@noop {} {\emph {\bibinfo {title} {Gas turbine combustion:
  alternative fuels and emissions}}}\ (\bibinfo  {publisher} {CRC press},\
  \bibinfo {year} {2010})\BibitemShut {NoStop}%
\bibitem [{\citenamefont {Simeoni}\ \emph {et~al.}(2010)\citenamefont
  {Simeoni}, \citenamefont {Bryk}, \citenamefont {Gorelli}, \citenamefont
  {Krisch}, \citenamefont {Ruocco}, \citenamefont {Santoro},\ and\
  \citenamefont {Scopigno}}]{widomline}%
  \BibitemOpen
  \bibfield  {author} {\bibinfo {author} {\bibfnamefont {G.~G.}\ \bibnamefont
  {Simeoni}}, \bibinfo {author} {\bibfnamefont {T.}~\bibnamefont {Bryk}},
  \bibinfo {author} {\bibfnamefont {F.~A.}\ \bibnamefont {Gorelli}}, \bibinfo
  {author} {\bibfnamefont {M.}~\bibnamefont {Krisch}}, \bibinfo {author}
  {\bibfnamefont {G.}~\bibnamefont {Ruocco}}, \bibinfo {author} {\bibfnamefont
  {M.}~\bibnamefont {Santoro}}, \ and\ \bibinfo {author} {\bibfnamefont
  {T.}~\bibnamefont {Scopigno}},\ }\bibfield  {title} {\enquote {\bibinfo
  {title} {The widom line as the crossover between liquid-like and gas-like
  behaviour in supercritical fluids},}\ }\href {\doibase 10.1038/nphys1683}
  {\bibfield  {journal} {\bibinfo  {journal} {Nature Physics}\ }\textbf
  {\bibinfo {volume} {6}},\ \bibinfo {pages} {503--507} (\bibinfo {year}
  {2010})}\BibitemShut {NoStop}%
\bibitem [{\citenamefont {Wang}\ \emph {et~al.}(2022)\citenamefont {Wang},
  \citenamefont {Zhao}, \citenamefont {Shu},\ and\ \citenamefont
  {Wei}}]{MDdrop}%
  \BibitemOpen
  \bibfield  {author} {\bibinfo {author} {\bibfnamefont {Z.}~\bibnamefont
  {Wang}}, \bibinfo {author} {\bibfnamefont {L.}~\bibnamefont {Zhao},
  \bibfnamefont {Wanhuiand~Zhou}}, \bibinfo {author} {\bibfnamefont
  {G.}~\bibnamefont {Shu}}, \ and\ \bibinfo {author} {\bibfnamefont
  {H.}~\bibnamefont {Wei}},\ }\bibfield  {title} {\enquote {\bibinfo {title} {A
  molecular dynamic study of evaporation/supercritical-transition
  inter-relationship and multicomponents interaction for alkane/alcohol
  droplets},}\ }\href {\doibase 10.1063/5.0078471} {\bibfield  {journal}
  {\bibinfo  {journal} {Physics of Fluids}\ }\textbf {\bibinfo {volume} {34}},\
  \bibinfo {pages} {022002} (\bibinfo {year} {2022})},\ \Eprint
  {http://arxiv.org/abs/https://doi.org/10.1063/5.0078471}
  {https://doi.org/10.1063/5.0078471} \BibitemShut {NoStop}%
\bibitem [{\citenamefont {Poblador-Ibanez}\ and\ \citenamefont
  {Sirignano}(2021)}]{interface}%
  \BibitemOpen
  \bibfield  {author} {\bibinfo {author} {\bibfnamefont {J.}~\bibnamefont
  {Poblador-Ibanez}}\ and\ \bibinfo {author} {\bibfnamefont {W.~A.}\
  \bibnamefont {Sirignano}},\ }\bibfield  {title} {\enquote {\bibinfo {title}
  {Liquid-jet instability at high pressures with real-fluid interface
  thermodynamics},}\ }\href {\doibase 10.1063/5.0055294} {\bibfield  {journal}
  {\bibinfo  {journal} {Physics of Fluids}\ }\textbf {\bibinfo {volume} {33}},\
  \bibinfo {pages} {083308} (\bibinfo {year} {2021})},\ \Eprint
  {http://arxiv.org/abs/https://doi.org/10.1063/5.0055294}
  {https://doi.org/10.1063/5.0055294} \BibitemShut {NoStop}%
\bibitem [{\citenamefont {Jofre}\ and\ \citenamefont
  {Urzay}(2021)}]{analytical}%
  \BibitemOpen
  \bibfield  {author} {\bibinfo {author} {\bibfnamefont {L.}~\bibnamefont
  {Jofre}}\ and\ \bibinfo {author} {\bibfnamefont {J.}~\bibnamefont {Urzay}},\
  }\bibfield  {title} {\enquote {\bibinfo {title} {Transcritical
  diffuse-interface hydrodynamics of propellants in high-pressure combustors of
  chemical propulsion systems},}\ }\href {\doibase
  https://doi.org/10.1016/j.pecs.2020.100877} {\bibfield  {journal} {\bibinfo
  {journal} {Progress in Energy and Combustion Science}\ }\textbf {\bibinfo
  {volume} {82}},\ \bibinfo {pages} {100877} (\bibinfo {year}
  {2021})}\BibitemShut {NoStop}%
\bibitem [{\citenamefont {Crua}, \citenamefont {Manin},\ and\ \citenamefont
  {Pickett}(2017)}]{experimental}%
  \BibitemOpen
  \bibfield  {author} {\bibinfo {author} {\bibfnamefont {C.}~\bibnamefont
  {Crua}}, \bibinfo {author} {\bibfnamefont {J.}~\bibnamefont {Manin}}, \ and\
  \bibinfo {author} {\bibfnamefont {L.~M.}\ \bibnamefont {Pickett}},\
  }\bibfield  {title} {\enquote {\bibinfo {title} {On the transcritical mixing
  of fuels at diesel engine conditions},}\ }\href {\doibase
  https://doi.org/10.1016/j.fuel.2017.06.091} {\bibfield  {journal} {\bibinfo
  {journal} {Fuel}\ }\textbf {\bibinfo {volume} {208}},\ \bibinfo {pages}
  {535--548} (\bibinfo {year} {2017})}\BibitemShut {NoStop}%
\bibitem [{\citenamefont {Xu}\ \emph {et~al.}(2021)\citenamefont {Xu},
  \citenamefont {Zhu}, \citenamefont {Jin}, \citenamefont {Guo},\ and\
  \citenamefont {Fan}}]{MD}%
  \BibitemOpen
  \bibfield  {author} {\bibinfo {author} {\bibfnamefont {B.}~\bibnamefont
  {Xu}}, \bibinfo {author} {\bibfnamefont {Y.}~\bibnamefont {Zhu}}, \bibinfo
  {author} {\bibfnamefont {H.}~\bibnamefont {Jin}}, \bibinfo {author}
  {\bibfnamefont {Y.}~\bibnamefont {Guo}}, \ and\ \bibinfo {author}
  {\bibfnamefont {J.}~\bibnamefont {Fan}},\ }\bibfield  {title} {\enquote
  {\bibinfo {title} {Transcritical transition of the fluid around the
  interface},}\ }\href {\doibase 10.1063/5.0067825} {\bibfield  {journal}
  {\bibinfo  {journal} {Physics of Fluids}\ }\textbf {\bibinfo {volume} {33}},\
  \bibinfo {pages} {122106} (\bibinfo {year} {2021})},\ \Eprint
  {http://arxiv.org/abs/https://doi.org/10.1063/5.0067825}
  {https://doi.org/10.1063/5.0067825} \BibitemShut {NoStop}%
\bibitem [{\citenamefont {Ma}, \citenamefont {Lv},\ and\ \citenamefont
  {Ihme}(2017)}]{MA_DF}%
  \BibitemOpen
  \bibfield  {author} {\bibinfo {author} {\bibfnamefont {P.~C.}\ \bibnamefont
  {Ma}}, \bibinfo {author} {\bibfnamefont {Y.}~\bibnamefont {Lv}}, \ and\
  \bibinfo {author} {\bibfnamefont {M.}~\bibnamefont {Ihme}},\ }\bibfield
  {title} {\enquote {\bibinfo {title} {An entropy-stable hybrid scheme for
  simulations of transcritical real-fluid flows},}\ }\href {\doibase
  https://doi.org/10.1016/j.jcp.2017.03.022} {\bibfield  {journal} {\bibinfo
  {journal} {Journal of Computational Physics}\ }\textbf {\bibinfo {volume}
  {340}},\ \bibinfo {pages} {330--357} (\bibinfo {year} {2017})}\BibitemShut
  {NoStop}%
\bibitem [{\citenamefont {Boyd}\ and\ \citenamefont
  {Jarrahbashi}(2021{\natexlab{a}})}]{drop_shock}%
  \BibitemOpen
  \bibfield  {author} {\bibinfo {author} {\bibfnamefont {B.}~\bibnamefont
  {Boyd}}\ and\ \bibinfo {author} {\bibfnamefont {D.}~\bibnamefont
  {Jarrahbashi}},\ }\bibfield  {title} {\enquote {\bibinfo {title} {Numerical
  study of the transcritical shock-droplet interaction},}\ }\href {\doibase
  10.1103/PhysRevFluids.6.113601} {\bibfield  {journal} {\bibinfo  {journal}
  {Phys. Rev. Fluids}\ }\textbf {\bibinfo {volume} {6}},\ \bibinfo {pages}
  {113601} (\bibinfo {year} {2021}{\natexlab{a}})}\BibitemShut {NoStop}%
\bibitem [{\citenamefont {Migliorino}\ \emph {et~al.}()\citenamefont
  {Migliorino}, \citenamefont {Chapelier}, \citenamefont {Scalo},\ and\
  \citenamefont {Lodato}}]{osc_pre_sp}%
  \BibitemOpen
  \bibfield  {author} {\bibinfo {author} {\bibfnamefont {M.~T.}\ \bibnamefont
  {Migliorino}}, \bibinfo {author} {\bibfnamefont {J.-B.}\ \bibnamefont
  {Chapelier}}, \bibinfo {author} {\bibfnamefont {C.}~\bibnamefont {Scalo}}, \
  and\ \bibinfo {author} {\bibfnamefont {G.}~\bibnamefont {Lodato}},\ }\enquote
  {\bibinfo {title} {Assessment of spurious numerical oscillations in
  high-order spectral difference solvers for supercritical flows},}\ in\ \href
  {\doibase 10.2514/6.2018-4273} {\emph {\bibinfo {booktitle} {2018 Fluid
  Dynamics Conference}}},\ \Eprint
  {http://arxiv.org/abs/https://arc.aiaa.org/doi/pdf/10.2514/6.2018-4273}
  {https://arc.aiaa.org/doi/pdf/10.2514/6.2018-4273} \BibitemShut {NoStop}%
\bibitem [{\citenamefont {Migliorino}\ and\ \citenamefont
  {Scalo}(2020)}]{thermoacoustic}%
  \BibitemOpen
  \bibfield  {author} {\bibinfo {author} {\bibfnamefont {M.~T.}\ \bibnamefont
  {Migliorino}}\ and\ \bibinfo {author} {\bibfnamefont {C.}~\bibnamefont
  {Scalo}},\ }\bibfield  {title} {\enquote {\bibinfo {title} {Real-fluid
  effects on standing-wave thermoacoustic instability},}\ }\href {\doibase
  10.1017/jfm.2019.856} {\bibfield  {journal} {\bibinfo  {journal} {Journal of
  Fluid Mechanics}\ }\textbf {\bibinfo {volume} {883}},\ \bibinfo {pages} {A23}
  (\bibinfo {year} {2020})}\BibitemShut {NoStop}%
\bibitem [{\citenamefont {Dobrev}\ \emph {et~al.}(2016)\citenamefont {Dobrev},
  \citenamefont {Kolev}, \citenamefont {Rieben},\ and\ \citenamefont
  {Tomov}}]{material_interfaces}%
  \BibitemOpen
  \bibfield  {author} {\bibinfo {author} {\bibfnamefont {V.~A.}\ \bibnamefont
  {Dobrev}}, \bibinfo {author} {\bibfnamefont {T.~V.}\ \bibnamefont {Kolev}},
  \bibinfo {author} {\bibfnamefont {R.~N.}\ \bibnamefont {Rieben}}, \ and\
  \bibinfo {author} {\bibfnamefont {V.~Z.}\ \bibnamefont {Tomov}},\ }\bibfield
  {title} {\enquote {\bibinfo {title} {Multi-material closure model for
  high-order finite element lagrangian hydrodynamics},}\ }\href {\doibase
  https://doi.org/10.1002/fld.4236} {\bibfield  {journal} {\bibinfo  {journal}
  {International Journal for Numerical Methods in Fluids}\ }\textbf {\bibinfo
  {volume} {82}},\ \bibinfo {pages} {689--706} (\bibinfo {year} {2016})},\
  \Eprint
  {http://arxiv.org/abs/https://onlinelibrary.wiley.com/doi/pdf/10.1002/fld.4236}
  {https://onlinelibrary.wiley.com/doi/pdf/10.1002/fld.4236} \BibitemShut
  {NoStop}%
\bibitem [{\citenamefont {Abgrall}(1996)}]{Multicomponent1}%
  \BibitemOpen
  \bibfield  {author} {\bibinfo {author} {\bibfnamefont {R.}~\bibnamefont
  {Abgrall}},\ }\bibfield  {title} {\enquote {\bibinfo {title} {How to prevent
  pressure oscillations in multicomponent flow calculations: A quasi
  conservative approach},}\ }\href {\doibase
  https://doi.org/10.1006/jcph.1996.0085} {\bibfield  {journal} {\bibinfo
  {journal} {Journal of Computational Physics}\ }\textbf {\bibinfo {volume}
  {125}},\ \bibinfo {pages} {150--160} (\bibinfo {year} {1996})}\BibitemShut
  {NoStop}%
\bibitem [{\citenamefont {Lapenna}(2018)}]{LMa1}%
  \BibitemOpen
  \bibfield  {author} {\bibinfo {author} {\bibfnamefont {P.~E.}\ \bibnamefont
  {Lapenna}},\ }\bibfield  {title} {\enquote {\bibinfo {title}
  {Characterization of pseudo-boiling in a transcritical nitrogen jet},}\
  }\href {\doibase 10.1063/1.5038674} {\bibfield  {journal} {\bibinfo
  {journal} {Physics of Fluids}\ }\textbf {\bibinfo {volume} {30}},\ \bibinfo
  {pages} {077106} (\bibinfo {year} {2018})},\ \Eprint
  {http://arxiv.org/abs/https://doi.org/10.1063/1.5038674}
  {https://doi.org/10.1063/1.5038674} \BibitemShut {NoStop}%
\bibitem [{\citenamefont {Abgrall}\ and\ \citenamefont {Karni}(2001)}]{DB_OR}%
  \BibitemOpen
  \bibfield  {author} {\bibinfo {author} {\bibfnamefont {R.}~\bibnamefont
  {Abgrall}}\ and\ \bibinfo {author} {\bibfnamefont {S.}~\bibnamefont
  {Karni}},\ }\bibfield  {title} {\enquote {\bibinfo {title} {Computations of
  compressible multifluids},}\ }\href {\doibase
  https://doi.org/10.1006/jcph.2000.6685} {\bibfield  {journal} {\bibinfo
  {journal} {Journal of Computational Physics}\ }\textbf {\bibinfo {volume}
  {169}},\ \bibinfo {pages} {594--623} (\bibinfo {year} {2001})}\BibitemShut
  {NoStop}%
\bibitem [{\citenamefont {Boyd}\ and\ \citenamefont
  {Jarrahbashi}(2021{\natexlab{b}})}]{DB_HB}%
  \BibitemOpen
  \bibfield  {author} {\bibinfo {author} {\bibfnamefont {B.}~\bibnamefont
  {Boyd}}\ and\ \bibinfo {author} {\bibfnamefont {D.}~\bibnamefont
  {Jarrahbashi}},\ }\bibfield  {title} {\enquote {\bibinfo {title} {A
  diffuse-interface method for reducing spurious pressure oscillations in
  multicomponent transcritical flow simulations},}\ }\href {\doibase
  https://doi.org/10.1016/j.compfluid.2021.104924} {\bibfield  {journal}
  {\bibinfo  {journal} {Computers \& Fluids}\ }\textbf {\bibinfo {volume}
  {222}},\ \bibinfo {pages} {104924} (\bibinfo {year}
  {2021}{\natexlab{b}})}\BibitemShut {NoStop}%
\bibitem [{\citenamefont {Rodriguez}\ \emph {et~al.}(2018)\citenamefont
  {Rodriguez}, \citenamefont {Vidal}, \citenamefont {Koukouvinis},
  \citenamefont {Gavaises},\ and\ \citenamefont {McHugh}}]{PC-SAFT}%
  \BibitemOpen
  \bibfield  {author} {\bibinfo {author} {\bibfnamefont {C.}~\bibnamefont
  {Rodriguez}}, \bibinfo {author} {\bibfnamefont {A.}~\bibnamefont {Vidal}},
  \bibinfo {author} {\bibfnamefont {P.}~\bibnamefont {Koukouvinis}}, \bibinfo
  {author} {\bibfnamefont {M.}~\bibnamefont {Gavaises}}, \ and\ \bibinfo
  {author} {\bibfnamefont {M.}~\bibnamefont {McHugh}},\ }\bibfield  {title}
  {\enquote {\bibinfo {title} {Simulation of transcritical fluid jets using the
  pc-saft eos},}\ }\href {\doibase https://doi.org/10.1016/j.jcp.2018.07.030}
  {\bibfield  {journal} {\bibinfo  {journal} {Journal of Computational
  Physics}\ }\textbf {\bibinfo {volume} {374}},\ \bibinfo {pages} {444--468}
  (\bibinfo {year} {2018})}\BibitemShut {NoStop}%
\bibitem [{\citenamefont {Kawai}, \citenamefont {Terashima},\ and\
  \citenamefont {Negishi}(2015)}]{press}%
  \BibitemOpen
  \bibfield  {author} {\bibinfo {author} {\bibfnamefont {S.}~\bibnamefont
  {Kawai}}, \bibinfo {author} {\bibfnamefont {H.}~\bibnamefont {Terashima}}, \
  and\ \bibinfo {author} {\bibfnamefont {H.}~\bibnamefont {Negishi}},\
  }\bibfield  {title} {\enquote {\bibinfo {title} {A robust and accurate
  numerical method for transcritical turbulent flows at supercritical pressure
  with an arbitrary equation of state},}\ }\href {\doibase
  https://doi.org/10.1016/j.jcp.2015.07.047} {\bibfield  {journal} {\bibinfo
  {journal} {Journal of Computational Physics}\ }\textbf {\bibinfo {volume}
  {300}},\ \bibinfo {pages} {116--135} (\bibinfo {year} {2015})}\BibitemShut
  {NoStop}%
\bibitem [{\citenamefont {Pantano}, \citenamefont {Saurel},\ and\ \citenamefont
  {Schmitt}(2017)}]{vanserw}%
  \BibitemOpen
  \bibfield  {author} {\bibinfo {author} {\bibfnamefont {C.}~\bibnamefont
  {Pantano}}, \bibinfo {author} {\bibfnamefont {R.}~\bibnamefont {Saurel}}, \
  and\ \bibinfo {author} {\bibfnamefont {T.}~\bibnamefont {Schmitt}},\
  }\bibfield  {title} {\enquote {\bibinfo {title} {An oscillation free
  shock-capturing method for compressible van der waals supercritical fluid
  flows},}\ }\href {\doibase https://doi.org/10.1016/j.jcp.2017.01.057}
  {\bibfield  {journal} {\bibinfo  {journal} {Journal of Computational
  Physics}\ }\textbf {\bibinfo {volume} {335}},\ \bibinfo {pages} {780--811}
  (\bibinfo {year} {2017})}\BibitemShut {NoStop}%
\bibitem [{\citenamefont {Toro}, \citenamefont {Castro},\ and\ \citenamefont
  {Lee}(2015)}]{ader}%
  \BibitemOpen
  \bibfield  {author} {\bibinfo {author} {\bibfnamefont {E.~F.}\ \bibnamefont
  {Toro}}, \bibinfo {author} {\bibfnamefont {C.~E.}\ \bibnamefont {Castro}}, \
  and\ \bibinfo {author} {\bibfnamefont {B.~J.}\ \bibnamefont {Lee}},\
  }\bibfield  {title} {\enquote {\bibinfo {title} {A novel numerical flux for
  the 3d euler equations with general equation of state},}\ }\href {\doibase
  https://doi.org/10.1016/j.jcp.2015.09.037} {\bibfield  {journal} {\bibinfo
  {journal} {Journal of Computational Physics}\ }\textbf {\bibinfo {volume}
  {303}},\ \bibinfo {pages} {80--94} (\bibinfo {year} {2015})}\BibitemShut
  {NoStop}%
\bibitem [{\citenamefont {Hou}\ and\ \citenamefont
  {LeFloch}(1994)}]{math_fund}%
  \BibitemOpen
  \bibfield  {author} {\bibinfo {author} {\bibfnamefont {T.~Y.}\ \bibnamefont
  {Hou}}\ and\ \bibinfo {author} {\bibfnamefont {P.~G.}\ \bibnamefont
  {LeFloch}},\ }\bibfield  {title} {\enquote {\bibinfo {title} {Why
  nonconservative schemes converge to wrong solutions: error analysis},}\
  }\href {\doibase https://doi.org/10.1090/S0025-5718-1994-1201068-0}
  {\bibfield  {journal} {\bibinfo  {journal} {Mathematics of computation}\
  }\textbf {\bibinfo {volume} {62}},\ \bibinfo {pages} {497--530} (\bibinfo
  {year} {1994})}\BibitemShut {NoStop}%
\bibitem [{\citenamefont {Toro}(1998)}]{ad_toro}%
  \BibitemOpen
  \bibfield  {author} {\bibinfo {author} {\bibfnamefont {E.~F.}\ \bibnamefont
  {Toro}},\ }\enquote {\bibinfo {title} {Primitive, conservative and adaptive
  schemes for hyperbolic conservation laws},}\ in\ \href@noop {} {\emph
  {\bibinfo {booktitle} {Numerical Methods for Wave Propagation: Selected
  Contributions from the Workshop held in Manchester, U.K., Containing the
  Harten Memorial Lecture}}},\ \bibinfo {editor} {edited by\ \bibinfo {editor}
  {\bibfnamefont {E.~F.}\ \bibnamefont {Toro}}\ and\ \bibinfo {editor}
  {\bibfnamefont {J.~F.}\ \bibnamefont {Clarke}}}\ (\bibinfo  {publisher}
  {Springer Netherlands},\ \bibinfo {address} {Dordrecht},\ \bibinfo {year}
  {1998})\ pp.\ \bibinfo {pages} {323--385}\BibitemShut {NoStop}%
\bibitem [{\citenamefont {Lee}\ \emph {et~al.}(2013)\citenamefont {Lee},
  \citenamefont {Toro}, \citenamefont {Castro},\ and\ \citenamefont
  {Nikiforakis}}]{ad_osher}%
  \BibitemOpen
  \bibfield  {author} {\bibinfo {author} {\bibfnamefont {B.~J.}\ \bibnamefont
  {Lee}}, \bibinfo {author} {\bibfnamefont {E.~F.}\ \bibnamefont {Toro}},
  \bibinfo {author} {\bibfnamefont {C.~E.}\ \bibnamefont {Castro}}, \ and\
  \bibinfo {author} {\bibfnamefont {N.}~\bibnamefont {Nikiforakis}},\
  }\bibfield  {title} {\enquote {\bibinfo {title} {Adaptive osher-type scheme
  for the euler equations with highly nonlinear equations of state},}\ }\href
  {\doibase https://doi.org/10.1016/j.jcp.2013.03.046} {\bibfield  {journal}
  {\bibinfo  {journal} {Journal of Computational Physics}\ }\textbf {\bibinfo
  {volume} {246}},\ \bibinfo {pages} {165--183} (\bibinfo {year}
  {2013})}\BibitemShut {NoStop}%
\bibitem [{\citenamefont {Fjelde}\ and\ \citenamefont {Karlsen}(2002)}]{ad1}%
  \BibitemOpen
  \bibfield  {author} {\bibinfo {author} {\bibfnamefont {K.~K.}\ \bibnamefont
  {Fjelde}}\ and\ \bibinfo {author} {\bibfnamefont {K.~H.}\ \bibnamefont
  {Karlsen}},\ }\bibfield  {title} {\enquote {\bibinfo {title} {High-resolution
  hybrid primitive–conservative upwind schemes for the drift flux model},}\
  }\href {\doibase https://doi.org/10.1016/S0045-7930(01)00041-X} {\bibfield
  {journal} {\bibinfo  {journal} {Computers \& Fluids}\ }\textbf {\bibinfo
  {volume} {31}},\ \bibinfo {pages} {335--367} (\bibinfo {year}
  {2002})}\BibitemShut {NoStop}%
\bibitem [{\citenamefont {Toro}(2013)}]{toro2013riemann}%
  \BibitemOpen
  \bibfield  {author} {\bibinfo {author} {\bibfnamefont {E.~F.}\ \bibnamefont
  {Toro}},\ }\href {\doibase https://doi.org/10.1007/b79761} {\emph {\bibinfo
  {title} {Riemann solvers and numerical methods for fluid dynamics: a
  practical introduction}}}\ (\bibinfo  {publisher} {Springer Science \&
  Business Media},\ \bibinfo {year} {2013})\BibitemShut {NoStop}%
\bibitem [{\citenamefont {Soave}(1972)}]{SRK}%
  \BibitemOpen
  \bibfield  {author} {\bibinfo {author} {\bibfnamefont {G.}~\bibnamefont
  {Soave}},\ }\bibfield  {title} {\enquote {\bibinfo {title} {Equilibrium
  constants from a modified redlich-kwong equation of state},}\ }\href
  {\doibase https://doi.org/10.1016/0009-2509(72)80096-4} {\bibfield  {journal}
  {\bibinfo  {journal} {Chemical Engineering Science}\ }\textbf {\bibinfo
  {volume} {27}},\ \bibinfo {pages} {1197--1203} (\bibinfo {year}
  {1972})}\BibitemShut {NoStop}%
\bibitem [{\citenamefont {Peng}\ and\ \citenamefont {Robinson}(1976)}]{PR}%
  \BibitemOpen
  \bibfield  {author} {\bibinfo {author} {\bibfnamefont {D.-Y.}\ \bibnamefont
  {Peng}}\ and\ \bibinfo {author} {\bibfnamefont {D.~B.}\ \bibnamefont
  {Robinson}},\ }\bibfield  {title} {\enquote {\bibinfo {title} {A new
  two-constant equation of state},}\ }\href {\doibase 10.1021/i160057a011}
  {\bibfield  {journal} {\bibinfo  {journal} {Industrial \& Engineering
  Chemistry Fundamentals}\ }\textbf {\bibinfo {volume} {15}},\ \bibinfo {pages}
  {59--64} (\bibinfo {year} {1976})},\ \Eprint
  {http://arxiv.org/abs/https://doi.org/10.1021/i160057a011}
  {https://doi.org/10.1021/i160057a011} \BibitemShut {NoStop}%
\bibitem [{\citenamefont {Poormahmood}\ and\ \citenamefont
  {Farshchi}(2020)}]{UC1}%
  \BibitemOpen
  \bibfield  {author} {\bibinfo {author} {\bibfnamefont {A.}~\bibnamefont
  {Poormahmood}}\ and\ \bibinfo {author} {\bibfnamefont {M.}~\bibnamefont
  {Farshchi}},\ }\bibfield  {title} {\enquote {\bibinfo {title} {Numerical
  study of the mixing dynamics of trans- and supercritical coaxial jets},}\
  }\href {\doibase 10.1063/5.0030183} {\bibfield  {journal} {\bibinfo
  {journal} {Physics of Fluids}\ }\textbf {\bibinfo {volume} {32}},\ \bibinfo
  {pages} {125105} (\bibinfo {year} {2020})},\ \Eprint
  {http://arxiv.org/abs/https://doi.org/10.1063/5.0030183}
  {https://doi.org/10.1063/5.0030183} \BibitemShut {NoStop}%
\bibitem [{\citenamefont {Harten}\ and\ \citenamefont
  {Hyman}(1983)}]{entropy_fix}%
  \BibitemOpen
  \bibfield  {author} {\bibinfo {author} {\bibfnamefont {A.}~\bibnamefont
  {Harten}}\ and\ \bibinfo {author} {\bibfnamefont {J.~M.}\ \bibnamefont
  {Hyman}},\ }\bibfield  {title} {\enquote {\bibinfo {title} {Self adjusting
  grid methods for one-dimensional hyperbolic conservation laws},}\ }\href
  {\doibase https://doi.org/10.1016/0021-9991(83)90066-9} {\bibfield  {journal}
  {\bibinfo  {journal} {Journal of Computational Physics}\ }\textbf {\bibinfo
  {volume} {50}},\ \bibinfo {pages} {235--269} (\bibinfo {year}
  {1983})}\BibitemShut {NoStop}%
\bibitem [{\citenamefont {Wang}\ and\ \citenamefont {Hickey}(2022)}]{StARS}%
  \BibitemOpen
  \bibfield  {author} {\bibinfo {author} {\bibfnamefont {J.~C.}\ \bibnamefont
  {Wang}}\ and\ \bibinfo {author} {\bibfnamefont {J.~P.}\ \bibnamefont
  {Hickey}},\ }\bibfield  {title} {\enquote {\bibinfo {title} {A class of
  structurally complete approximate riemann solvers for trans-and supercritical
  flows with large gradients},}\ }\href@noop {} {\bibfield  {journal} {\bibinfo
   {journal} {Available at SSRN 4010625}\ } (\bibinfo {year}
  {2022})}\BibitemShut {NoStop}%
\bibitem [{\citenamefont {Wang}\ and\ \citenamefont
  {Hickey}(2020)}]{expansion}%
  \BibitemOpen
  \bibfield  {author} {\bibinfo {author} {\bibfnamefont {J.~C.~H.}\
  \bibnamefont {Wang}}\ and\ \bibinfo {author} {\bibfnamefont {J.-P.}\
  \bibnamefont {Hickey}},\ }\bibfield  {title} {\enquote {\bibinfo {title}
  {Analytical solutions to shock and expansion waves for non-ideal equations of
  state},}\ }\href {\doibase 10.1063/5.0015531} {\bibfield  {journal} {\bibinfo
   {journal} {Physics of Fluids}\ }\textbf {\bibinfo {volume} {32}},\ \bibinfo
  {pages} {086105} (\bibinfo {year} {2020})},\ \Eprint
  {http://arxiv.org/abs/https://doi.org/10.1063/5.0015531}
  {https://doi.org/10.1063/5.0015531} \BibitemShut {NoStop}%
\bibitem [{\citenamefont {Menikoff}\ and\ \citenamefont
  {Plohr}(1989)}]{noncv_EOS}%
  \BibitemOpen
  \bibfield  {author} {\bibinfo {author} {\bibfnamefont {R.}~\bibnamefont
  {Menikoff}}\ and\ \bibinfo {author} {\bibfnamefont {B.~J.}\ \bibnamefont
  {Plohr}},\ }\bibfield  {title} {\enquote {\bibinfo {title} {The riemann
  problem for fluid flow of real materials},}\ }\href {\doibase
  10.1103/RevModPhys.61.75} {\bibfield  {journal} {\bibinfo  {journal} {Rev.
  Mod. Phys.}\ }\textbf {\bibinfo {volume} {61}},\ \bibinfo {pages} {75--130}
  (\bibinfo {year} {1989})}\BibitemShut {NoStop}%
\bibitem [{\citenamefont {M{\"u}ller}\ and\ \citenamefont
  {Vo{\ss}}(1999)}]{muller1999riemann}%
  \BibitemOpen
  \bibfield  {author} {\bibinfo {author} {\bibfnamefont {S.}~\bibnamefont
  {M{\"u}ller}}\ and\ \bibinfo {author} {\bibfnamefont {A.}~\bibnamefont
  {Vo{\ss}}},\ }\bibfield  {title} {\enquote {\bibinfo {title} {A riemann
  solver for the euler equations with non-convex equation of state},}\
  }\href@noop {} {\  (\bibinfo {year} {1999})}\BibitemShut {NoStop}%
\bibitem [{\citenamefont {GUARDONE}(2007)}]{real_Roe}%
  \BibitemOpen
  \bibfield  {author} {\bibinfo {author} {\bibfnamefont {A.}~\bibnamefont
  {GUARDONE}},\ }\bibfield  {title} {\enquote {\bibinfo {title}
  {Three-dimensional shock tube flows for dense gases},}\ }\href {\doibase
  10.1017/S0022112007006313} {\bibfield  {journal} {\bibinfo  {journal}
  {Journal of Fluid Mechanics}\ }\textbf {\bibinfo {volume} {583}},\ \bibinfo
  {pages} {423–442} (\bibinfo {year} {2007})}\BibitemShut {NoStop}%
\bibitem [{\citenamefont {Kim}, \citenamefont {Choi},\ and\ \citenamefont
  {Kim}(2012)}]{KIM}%
  \BibitemOpen
  \bibfield  {author} {\bibinfo {author} {\bibfnamefont {S.-K.}\ \bibnamefont
  {Kim}}, \bibinfo {author} {\bibfnamefont {H.-S.}\ \bibnamefont {Choi}}, \
  and\ \bibinfo {author} {\bibfnamefont {Y.}~\bibnamefont {Kim}},\ }\bibfield
  {title} {\enquote {\bibinfo {title} {Thermodynamic modeling based on a
  generalized cubic equation of state for kerosene/lox rocket combustion},}\
  }\href {\doibase https://doi.org/10.1016/j.combustflame.2011.10.008}
  {\bibfield  {journal} {\bibinfo  {journal} {Combustion and Flame}\ }\textbf
  {\bibinfo {volume} {159}},\ \bibinfo {pages} {1351--1365} (\bibinfo {year}
  {2012})}\BibitemShut {NoStop}%
\bibitem [{\citenamefont {Dumbser}\ and\ \citenamefont {Toro}(2011)}]{dot}%
  \BibitemOpen
  \bibfield  {author} {\bibinfo {author} {\bibfnamefont {M.}~\bibnamefont
  {Dumbser}}\ and\ \bibinfo {author} {\bibfnamefont {E.~F.}\ \bibnamefont
  {Toro}},\ }\bibfield  {title} {\enquote {\bibinfo {title} {A simple extension
  of the osher riemann solver to non-conservative hyperbolic systems},}\ }\href
  {\doibase https://doi.org/10.1007/s10915-010-9400-3} {\bibfield  {journal}
  {\bibinfo  {journal} {Journal of Scientific Computing}\ }\textbf {\bibinfo
  {volume} {48}},\ \bibinfo {pages} {70--88} (\bibinfo {year}
  {2011})}\BibitemShut {NoStop}%
\bibitem [{\citenamefont {Arabi}, \citenamefont {Trépanier},\ and\
  \citenamefont {Camarero}(2017)}]{extension_Roe}%
  \BibitemOpen
  \bibfield  {author} {\bibinfo {author} {\bibfnamefont {S.}~\bibnamefont
  {Arabi}}, \bibinfo {author} {\bibfnamefont {J.-Y.}\ \bibnamefont
  {Trépanier}}, \ and\ \bibinfo {author} {\bibfnamefont {R.}~\bibnamefont
  {Camarero}},\ }\bibfield  {title} {\enquote {\bibinfo {title} {A simple
  extension of roe's scheme for real gases},}\ }\href {\doibase
  https://doi.org/10.1016/j.jcp.2016.10.067} {\bibfield  {journal} {\bibinfo
  {journal} {Journal of Computational Physics}\ }\textbf {\bibinfo {volume}
  {329}},\ \bibinfo {pages} {16--28} (\bibinfo {year} {2017})}\BibitemShut
  {NoStop}%
\end{thebibliography}%

\end{document}